\documentstyle[preprint,aps,prd]{revtex}
\draft
\begin{document}
\title{
\begin{flushright}{\normalsize Saclay--T96/133 \\}
\end{flushright}
Damping rates of hard momentum particles\\ in a cold ultrarelativistic plasma}
\date{\today}
\author{Beno\^\i t Vanderheyden}
\address{Department of Physics, University of Illinois at
        Urbana-Champaign, Urbana, IL 61801}
\author{Jean-Yves Ollitrault\footnote{Affiliated with CNRS}}
\address{Service
de Physique Th\'eorique\footnote{Laboratoire de la Direction des Sciences
de la Mati\`ere du Commissariat \`a l'Energie Atomique}, CE-Saclay, 
91191 Gif-sur-Yvette cedex, France}
\maketitle

\begin{abstract} 
We compute the damping rates of one--particle excitations in a 
cold ultrarelativistic plasma to leading order in the 
coupling constant $e$ for three types of interaction: 
Yukawa coupling to a massless scalar boson, QED and QCD. 
Damping rates of charged particles in QED and QCD are of order $e^3\mu$, 
while damping rates of other particles are of order $e^4\mu$ or 
$e^4\mu\log(1/e)$. 
We find that the damping rate of an electron or of a quark 
is constant far from the Fermi surface, 
and decreases linearly with the excitation energy close 
to the Fermi surface. This unusual behavior is attributed to
the long--range magnetic interactions. 
\end{abstract}

\newpage
\section{Introduction}

The quasiparticle concept is a powerful tool for studying the 
dynamical properties of ultrarelativistic plasmas; it has been 
widely  used in the recent litterature~\cite{reviews}.  
Weakly excited states of a plasma can be described as superpositions of 
quasiparticle states, which behave, at least in a first approximation, 
as free, non--interacting, particles. These elementary excitations 
undergo damping through their mutual coupling, which gives their 
energy spectrum a finite width. 
The quasiparticle concept holds only if this energy 
spread, or equivalently the damping rate, 
is negligible compared to the mean energy of their wave packet.

In this paper, we compute damping rates in plasmas at high density 
and zero temperature, $T=0$. 
While the thermodynamic properties of ultrarelativistic 
degenerate plasmas are well known~\cite{Freedman},
little work has been devoted so far to their dynamical 
properties~\cite{BlaiOll,finitemu}. 
Most works on damping rates have been focusing on plasmas at high 
temperature with vanishing chemical potential, $\mu=0$.
However, plasmas with finite chemical potential are also relevant
in view of phenomenological applications: 
first, degenerate quark systems (with $\mu\ll T$) might exist in the cores of 
neutron stars~\cite{pulsar}; second, the state of the hadronic matter
(possibly a quark--gluon plasma) 
temporarily formed in an ultrarelativistic nucleus--nucleus collision 
corresponds to values of $\mu$ and $T$ of the same order of magnitude 
at the presently available energies~\cite{Stachel}.

   From a theoretical point of view, the zero temperature case owes its
interest to the basic difference, compared to the high temperature limit,
in the infrared divergences which plague perturbative 
calculations in gauge theories. 
Naive perturbation theory leads to two levels of 
infrared divergences~\cite{Braaten94,Rebhan95}. 
Those appearing in the calculation of transport coefficients~\cite{BayMon} and 
collisional energy loss~\cite{BraTho} are cured by taking into account 
screening effects at the one--loop level: 
propagators and vertices must be corrected at long wavelengths, 
according to the resummation scheme developed by Braaten and 
Pisarski~\cite{BraPis}. 
The momentum and energy dependence of these medium effects is 
the same for hot ($\mu=0$) and cold ($T=0$) relativistic plasmas, 
and is characterized by a scale of order $eT$ for a hot plasma, 
and $e\mu$ for a cold plasma. 
At high temperature, the lack of static screening
of the transverse part of the interaction is responsible for 
a logarithmic divergence in the perturbative calculation of 
the fermion damping rate~\cite{Pisarski89,Lebedev90,Pisarski93}. 
This problem has been recently solved in hot QED~\cite{BlaIan96}. 
By contrast, this second level of divergence does not appear at 
zero temperature, because of Pauli blocking. 
This allows us to obtain finite, most often analytical 
expressions for the damping rates $\Gamma(p)$ of all one-particle 
excitations, as a function of their momentum $p$. 

Three types of interactions are discussed and compared. 
(1) Electromagnetic interaction (QED): this corresponds to a degenerate 
electron gas.
(2) Non-abelian $SU(N_c)$ interaction (QCD): for $N_c=3$, this corresponds 
to degenerate quark matter. 
(3) In addition, we consider a toy model where the fermion field $\psi(x)$
is coupled to a massless scalar field $\phi(x)$, with the Yukawa interaction 
${\cal L}_Y=e\bar\psi\psi\phi$. 
These theories share several features. 
The resummed fermion propagator is the same for the three theories. 
The resummed boson propagator is the same for the two gauge theories, 
QED and QCD, while screening corrections reduce to a medium--induced 
mass term~\cite{Thoma95} for the Yukawa interaction. 
The coupling constant, noted $e$ for all theories, will 
be assumed to be much smaller than unity, $e\ll 1$, to ensure that a 
perturbative expansion is reliable.
It has been shown that perturbation theory can be used to study the properties 
of the QCD phase if the temperature and/or the chemical potential is much 
larger than $\Lambda_{QCD}$~\cite{Freedman}. 

We are considering degenerate Fermi systems in their ground state. 
Single particle excitations are obtained by adding or removing
one particle from the ground state, and
damping of these excitations results from collisions with the fermions of the 
Fermi sea.  
Two types of excitations compose the quasiparticle spectrum. At large
(``hard'') momenta $p$, $p \sim \mu \gg e\mu$, the elementary modes 
correspond to single particles slightly perturbed by the medium.
Their damping processes are the object of this paper.
For hard fermion or hole excitations, the only collisional process 
to leading order in $e$ is elastic (M\o ller) scattering $ff\rightarrow ff$. 
[We use 
the generic names bosons ($b$), fermions ($f$) and antifermions ($\bar f$).] 
For an antifermion, two processes may contribute, namely elastic 
(Bhabha) scattering $\bar ff\rightarrow \bar ff$ and pair annihilation 
$\bar ff\rightarrow bb$. 
For a boson, the only process is elastic (Compton) scattering 
$bf\rightarrow bf$. 
On the other hand, long wavelength (``soft'') modes, of momentum $p$ of
order $e \mu$, correspond to collective oscillations in the medium.
Their damping processes 
are more complex: to leading order in $e$, bremsstrahlung contributes: 
$ff\rightarrow ffb$ for a soft hole, or 
$\bar fff\rightarrow bf$ for a soft antifermion. 
A discussion of these calculations will be presented in a forthcoming 
publication\cite{VO}.
 
Depending on the interaction and on the type of excitation considered, 
the dominant contribution to the damping rate may come either 
from processes with large scattering angles ($\theta\sim 1$), 
from processes with small scattering angles ($\theta\sim e$), or from both. 
Large scattering angles correspond to hard exchanged quanta,  
for which medium effects are small: then, the amplitudes
are the same as in the vacuum. 
On the other hand, if a hard particle of 
momentum $p\sim\mu$ is deflected by an angle of order $e$, 
the exchanged quantum has a soft momentum of order $e\mu$, 
for which medium effects become important. 
In section~\ref{s:general}, we estimate the orders of magnitude of both 
hard and soft contributions to the damping rates. We show how they 
can be calculated, in a kinetic approach, 
as phase space integrals of elementary scattering processes. 
The equivalence with field theoretical methods is 
recalled in Appendix B.
Section~\ref{s:body} presents the calculations of damping rates for the 
various excitations considered, while section~\ref{s:discussion} is devoted to
a discussion of the results along with a comparison between damping rates in 
cold and hot plasmas.

\section{General analysis}
\label{s:general}

\subsection{Orders of magnitude}
\label{s:magnitude}

The damping rate $\Gamma$, the number of collisions per unit time, is of order 
$\Gamma\sim \sigma nv$ where $\sigma$ is the scattering cross section, 
$n$ the density of scatterers and $v$ the relative velocity. 
 In this paper, we assume that all particles are massless,
which implies $v=1$ ($\hbar=c=1$). 
We consider $2\rightarrow 2$ elementary processes in which 
an incoming particle of four-momentum $P=(p,\hbox{\boldmath$p$})$ 
(fermion above the Fermi level, antifermion or boson) is added to the system 
and scatters on a particle of the Fermi sea  
with four-momentum $K=(k,\hbox{\boldmath$k$})$, $k<\mu$.  
We denote by $P'=(p',\hbox{\boldmath$p'$})$ and 
$K'=(k',\hbox{\boldmath$k'$})$ the four-momenta of outgoing particles. 
If the excitation under study is a hole in the Fermi sea, incoming and 
outgoing particles must be interchanged. 
The differential cross section for unpolarized particles 
can generally be written as 
\begin{equation}
\label{sigma}
{d\sigma\over dt}={1\over 32\pi s^2}\overline{|M|^2},
\end{equation}
where $\overline{|M|^2}$ denotes the scattering matrix element squared, 
averaged over the helicity states of the incoming particle with momentum 
\hbox{\boldmath$p$},
and summed over the helicity states of the other particles. 
(Note that we do a sum, rather than an average, over the 2 helicity states 
of the scatterer. This unusual convention will turn out to be 
convenient in the following sections.) 
A factor $1/2$ is included in $\overline{|M|^2}$ if the outgoing particles 
are identical. 
The variables $s=(P+K)^2$, $t=(P-P')^2$ and $u=(P-K')^2=-s-t$ 
are the usual Mandelstam variables. 
In the center of mass frame, $t$ 
is related to the momentum transfer $q$ by $t=-q^2$ and to 
the scattering angle $\theta$ by $t=-s\sin^2(\theta/2)$. 
In order to obtain the total cross section, one must 
integrate eq.~(\ref{sigma}) over $t$ between $t=-s$ and $t=0$.

For a given scattering process in the plasma, 
the squared matrix element $\overline{|M|^2}$ is deduced from the usual 
Feynman rules with appropriate  corrections taking
into account medium polarization effects. We distinguish three cases, 
depending on the behaviour of $\overline{|M|^2}$ at small $t$:

1) The tree matrix element $\overline{|M|^2}$ is finite at small momentum 
transfers: one 
example is fermion--fermion scattering in a Yukawa theory, for which 
the tree matrix element  is
\begin{equation}
\label{Melyu}
\overline{|M|^2}=3  e^4.
\end{equation}
The total cross section is therefore finite and of order
$\sigma\sim e^4/\mu^2$, since $s$ is of order $\mu^2$. 
The density of particles per unit volume in 
the Fermi sea is of order $n\sim\mu^3$, which gives $\Gamma\sim e^4\mu$. 

2) $\overline{|M|^2}$ is proportional to $1/t^2$ at small $t$: 
the differential cross section grows like $1/\theta^4$ at small scattering
angle $\theta$, which makes the total cross section diverge. This 
occurs in collisions between charged particles (Rutherford divergence).
For electron--electron scattering in QED (M\o ller scattering),
\begin{equation}
\overline{|M|^2}=4  e^4\left(\frac{u}{t}+\frac{t}{u}+1\right)^2, 
\label{MelQED}
\end{equation}
so that the total cross section diverges linearly in the infrared: 
$d\sigma\sim e^4 dt/t^2$.
However, in a plasma, due to the screening of
the electric charge, the interaction potential decreases exponentially 
for distances larger than the Debye screening length $r_D\sim 1/(e\mu)$. 
This causes a saturation of the differential cross section for momentum 
transfers $q<r_D^{-1}\sim e\mu$, i.e. for scattering angles 
$\theta<e$, or equivalently for $t<t_{\min}=e^2\mu^2$. 
The total cross section is therefore
of order $\sigma\sim e^4/t_{\min}\sim  e^2/\mu^2$. 
Note that this argument is not completely valid for the transverse part 
of the interaction, which is screened dynamically only (i.e. at finite energy
transfer). However, as we shall show in detail later, dynamical screening is
sufficient at zero temperature to saturate the damping rate at small momentum
transfer.
Now, for small momentum transfers of order $ e\mu$, 
the density of scatterers is 
no longer $n\sim\mu^3$:
because of Pauli blocking, the outgoing electron must have 
an energy larger than the Fermi energy $\mu$, and 
only the electrons just below the Fermi surface can contribute, within 
an interval of order $ e\mu$. Their density is $n\sim e\mu^3$, and one 
finally obtains $\Gamma\sim\sigma n\sim e^3\mu$. 

3) $\overline{|M|^2}$ is proportional to $1/t$ at small $t$. 
This happens in processes where the exchanged particle is a massless fermion, 
such as pair annihilation and Compton scattering. 
For instance, in the Yukawa theory, the square tree matrix element for 
pair annihilation into two massless Yukawa bosons ($f\bar f\to b b$) is 
\begin{eqnarray}
\label{melannyu}
\overline{|M|^2}&=& { e^4\over 2} \left(\frac{u}{t}+\frac{t}{u}-2\right),
\end{eqnarray}
and the total cross section diverges logarithmically in the infrared, 
rather than linearly in the previous case. 
This divergence is also cured by collective effects, which 
become important when the momentum transfer is of order $e\mu$. 
Note that these medium effects are taken into account in the fermion 
propagator, instead of the boson propagator in the previous case. 
Following the same reasoning, the 
total cross section is thus of order $\sigma= e^4\ln(1/e)/\mu^2$.
For an annihilation process, there is no Pauli blocking in the 
final state, and the density of scatterers is of order $n\sim\mu^3$,  
hence the damping rate is of order $\Gamma\sim e^4\ln(1/e)\mu$.
In the case of Compton scattering, this last remark applies
only for a boson whose energy is larger than the Fermi energy, 
so that it can turn into a fermion with almost the same 
momentum. Otherwise, Pauli blocking alone inhibits the infrared 
divergence and the damping rate is of order $ e^4\mu$ (section~\ref{s:boson}).

These considerations on orders of magnitude allow us to simplify the 
damping rate calculation. In case 1),
kinematics is dominated by hard momentum transfer (the ``hard sector 
contribution''): the phase space for soft momentum transfer 
(the ``soft sector contribution'') is smaller by
a factor $e$. On the other hand, soft momentum transfer dominates in case 2);
the differential cross  section integrated over momentum transfer of order 
$e\mu$ is larger by a factor $1/e^2$ than the contribution
from hard momentum transfer, while the phase space is smaller by 
only a factor $e$.
Finally, in case 3), both hard and soft sector contributions
are of equal magnitude. 

We separate the hard and soft scales 
by introducing an arbitrary IR (resp. UV) momentum cutoff 
$q_*$ in the hard (resp. soft) contribution. The cutoff $q_*$ is chosen at
an intermediate scale, $ e\mu\ll q_*\ll\mu$. 
The hard and soft sector contributions to the damping rate are both 
computed in the framework of kinetic theory. 
For $q>q_*$ (hard contribution), collective effects are negligible 
since $q\gg e\mu$, and the damping rate can be calculated from
the vacuum scattering amplitudes, 
integrated over phase space with an appropriate choice of variables, described
in section~\ref{s:hard contribution}. 
For $q<q_*$ (soft contribution), medium effects must be taken into account,
but the kinematics is simplified by the fact that the exchanged particle 
is much softer than the external ones.
The damping rate is 
most simply calculated as the emission probability of a soft, off--shell
quantum, as developed in section~\ref{s:soft contribution}. 
In case 3), both approximations are compatible in the region of the cutoff, 
so that the total damping rate, which is the sum of the two  
contributions, does not depend on the cutoff $q_*$  
\cite{BraYuan}.

Note that processes involving soft external particles, either
incoming or outgoing, are subleading because the associated 
phase space is small. Thus, the external legs in the processes 
considered in this paper will always be hard. The situation 
is different, of course, when studying the damping of soft excitations 
\cite{VO}, where at least the incoming particle is soft.

\subsection{Hard sector contribution}
\label{s:hard contribution}

The hard sector contribution is calculated as the  
total transition rate, integrated over the available phase space:
\begin{eqnarray}
\label{Gammaex1}
\Gamma_h(p)={1\over 2 p}\int{d\tau_{p'}d\tau_k d\tau_{k'}\overline{|M|^2}
(2\pi)^4\delta^4(P+K-P'-K')}.
\end{eqnarray}
Here, $d\tau_k=d^3 k/2k (2\pi)^3$ is the Lorentz invariant phase space volume.
When all the particles involved in the scattering process are hard, including 
the exchanged one, medium effects can be ignored to leading order, and the 
matrix element is computed with the usual Feynman rules. 
If the outgoing particles are electrons, the phase space is 
limited by the Pauli exclusion principle: their energy must 
be larger than the Fermi energy $\mu$.

Alternatively, we could have chosen to compute the damping rate in 
a field theoretical approach, as the imaginary part of the self-energy. 
Generally, the leading hard sector contribution corresponds 
to the imaginary part of two-loop energy diagrams, as illustrated in 
Fig.~1 in the case of electron--electron scattering. 
This equivalence is proven explicitly in Appendix~B. 

The phase space integration in eq.~(\ref{Gammaex1}) is easily carried out 
if an appropriate choice of kinematic variables is made: there are 9 
integration variables, and 4 constraints, 
hence 5 degrees of freedom. Since the transition probability
is invariant under a 
simultaneous rotation of $\hbox{\boldmath$p'$}$, $\hbox{\boldmath$k$}$ 
and $\hbox{\boldmath$k'$}$ about the direction of 
$\hbox{\boldmath$p$}$, by integrating over this 
angle only 4 degrees of freedom remain.  
At this point, it is convenient to introduce the 4-momentum 
$Q=(\omega,\hbox{\boldmath$q$})$ transferred by the incoming particle to the 
plasma:
\begin{equation}
\label{defwq}
Q=(\omega,\hbox{\boldmath$q$})\equiv P-P'=K'-K
\end{equation} 
We choose 3 variables as $k$,	$\omega$ and $q\equiv|\hbox{\boldmath$q$}|$. 
Note that $k$ and $\omega$ fix the energy of the incoming and 
outgoing particles, while $q$ fixes the angle between 
$\hbox{\boldmath$k$}$ and $\hbox{\boldmath$k'$}$. 
The fourth degree of freedom corresponds to the angle between the 
plane spanned by $(\hbox{\boldmath$p$},\hbox{\boldmath$p'$})$ and the plane 
spanned by $(\hbox{\boldmath$k$},\hbox{\boldmath$k'$})$ 
or, equivalently, to the azimuthal angle $\phi$ of $\hbox{\boldmath$k$}$ 
around $\hbox{\boldmath$q$}$. With these variables, eq.~(\ref{Gammaex1}) 
becomes
\begin{equation}
\label{Gammaex11}
\Gamma_h(p)={1\over 128\pi^3 p^2}\int{dk\, d\omega\, dq
\left<\overline{|M|^2}\right>}
\end{equation}
where the brackets $\left<\ \right>$ denote an average over the 
azimuthal angle $\phi$. 

The integration limits on $q$ are easily derived from the definitions in
eq.~(\ref{defwq}): 
\begin{equation}
\label{qlimit}
|\omega| < q < \min(k'+k,p'+p) = \min (2k+\omega, 2p-\omega).
\end{equation}
Since the scattered fermion belongs to the Fermi sea, we have the 
constraint $k<\mu$. 
Depending on whether or not the outgoing particles are fermions, 
additional constraints on $k$ and $\omega$ may result from the 
Pauli blocking conditions: $k'=k+\omega>\mu$ and/or $p'=p-\omega>\mu$. 
Finally, the momentum transfer is kept hard by imposing $q>q_*$ with
$e \mu \ll q_* \ll \mu$.
Together with eq.~(\ref{qlimit}), these conditions completely specify 
the integration domain in eq.~(\ref{Gammaex11}).

\subsection{Soft sector contribution}
\label{s:soft contribution}

When the momentum transfer of a scattering process 
is of order $e\mu$, one must correct the bare interaction for medium 
(screening) effects, whose contribution in the propagator is of the same 
order as the bare propagator itself. 
This is achieved by replacing the bare propagator by 
the resummed propagator given in Appendix A. 
The damping rate is then evaluated as a collision integral similar 
to eq.~(\ref{Gammaex1}), however with a screened interaction, as illustrated 
in Fig.~2 (left) for electron--electron scattering. 

The soft contribution to the damping rate can in fact be cast into a much 
simpler form, as the transition rate of a process where the 
the incoming hard particle, with four-momentum $P=(p,\hbox{\boldmath$p$})$, 
emits a virtual (spacelike) soft particle with four-momentum 
$Q=(\omega,\hbox{\boldmath$q$})$ and scatters into a 
hard particle with four-momentum $P'=(p',\hbox{\boldmath$p'$})$ 
(Fig.~2, right). 
The fact that the soft particle is spacelike
means that the actual physical process is, in the example of Fig.~2, 
electron--electron scattering, not ``electron decay''. 
We show in Appendix B that the two approaches (Fig.~2, left, and Fig.~2, 
right) are equivalent and 
amount to evaluate the imaginary part of a (resummed) one--loop self energy 
diagram (Fig.~2, middle). 

The damping rate $\Gamma(p)$, for the process in Fig.~2 (right), can 
be evaluated from Fermi's Golden 
Rule, as in eq.~(\ref{Gammaex1}), with an important modification: 
the soft energy $\omega$ and momentum $q=|\hbox{\boldmath$q$}|$ are 
no longer related 
by a dispersion relation. Instead, a whole range of values are allowed 
according to a spectral distribution whose actual form is derived from the 
screening corrections.
The damping rate takes then the following form:
\begin{eqnarray}
\label{gammasoft}
\displaystyle\Gamma_s(p)
&=&{1\over 2p}\int{
{d^4 Q\over (2\pi)^4}\,\rho(Q)
\,2\pi\delta((P-Q)^2)}\, \overline{|M|^2}.
\end{eqnarray}
In this expression, $\rho(Q)$ denotes the spectral function of the
soft particle. The spectral function of the outgoing hard particle 
is simply $2\pi\delta((P-Q)^2)$. $\overline{|M|^2}$ 
denotes the matrix element of the transition on the right of  
Fig.~2, squared and summed over final spins, and averaged over the 
spin states of the incoming particle.

In the limit where $q$ is much softer than $p$, 
$(P-Q)^2=-2p(\omega-q\cos\theta)$, where $\theta$ is 
the angle between $\hbox{\boldmath$q$}$ and $\hbox{\boldmath$p$}$. 
Thus the condition that the 
outgoing hard particle is on its mass shell, $(P-Q)^2=0$ reduces to 
$\cos\theta=\omega/q$. Integrating over $\theta$, the previous equation 
becomes 
\begin{equation}
\label{gammasoftp}
\Gamma_s(p)={1\over 16\pi^2p^2}\int_{0}^{q_*} qdq\int_{-q}^{q}d\omega
\,\rho(\omega,q)\overline{|M|^2}.
\end{equation}
Note that $\omega>0$ (resp. $\omega<0$) corresponds to the 
emission (resp. absorption) of a soft particle.

The phase space is further restricted by the Fermi--Dirac and Bose--Einstein 
distributions, eqs.~(\ref{BEFD}). 
If the outgoing hard particle is a fermion,
it must be above the Fermi level, which implies $\omega<p-\mu$. 
If the soft particle is a boson, the only possibility 
is $\omega>0$ because there is no boson initially present at $T=0$. 
If the soft particle is a fermion, both signs are possible:
$\omega<0$ corresponds to the absorption of a soft fermion, and 
$\omega>0$ to the emission of a soft antifermion.
Note that the phase space for the soft virtual particle 
is limited by the same statistical constraints as if it were a real, 
on--shell, particle. 

It turns out that the matrix element squared $\overline{|M|^2}$ 
always takes a very simple form, as we shall see in the next section. 

\section{Results}
\label{s:body}

We now compute explicitly the damping rates of one--particle 
excitations to leading order in $e$ for the three theories 
(Yukawa, QED and QCD). We first study the fermionic excitations, 
in sections~\ref{s:fermion yuka} to \ref{s:fermion antifermion},
then the bosonic excitations, in sections~\ref{s:boson} and \ref{s:gluon}. 
For each type of excitation, we show 
the self--energy diagrams whose imaginary part 
corresponds to the elementary process under consideration.

\subsection{Fermion in Yukawa's theory}
\label{s:fermion yuka}

We consider an incoming fermion of momentum $\hbox{\boldmath$p$}$ above the 
Fermi level: $p \geq \mu$. 
In the Yukawa theory, the fermion--fermion scattering matrix element squared,
corresponding to the Feynman diagrams depicted in Fig.~3, is  a constant
given by eq.~(\ref{Melyu}). 
Collisions are thus dominated by hard momentum transfers, as discussed 
in section~\ref{s:magnitude}. 
The rate $\Gamma$ is given by eq.~(\ref{Gammaex11}), 
where Fermi statistics imposes the conditions
\begin{equation}
\label{contrainteel}
0<\mu-k < \omega < p-\mu.
\end{equation}
Using eq.~(\ref{contrainteel}), the 
limits on $q$ given by eqs.~(\ref{qlimit}) reduce to 
$\omega<q<2k+\omega$. 
The integration is straightforward and leads to the expressions
\begin{eqnarray}
\Gamma(p)=\left\{\matrix{\strut
\displaystyle\frac{e^4}{128 \pi^3}\ 
\frac{\mu^2 (3p-4\mu)}{p^2}\hfill &{\rm for}\ p > 2 \mu,\hfill \cr
\displaystyle\frac{e^4}{128 \pi^3}\ \frac{(p-\mu)^2\ (4\mu-p)}{p^2} 
&{\rm for}\ \mu < p < 2 \mu.\strut \cr
}\right. 
\label{Gelyu>b}
\end{eqnarray}
This damping rate corresponds to the imaginary part of the two--loop
diagrams displayed in Fig.~4. The third two--loop diagram, the 
rainbow diagram (see Fig.~7) does not contribute at zero temperature: 
its imaginary part corresponds to Compton scattering  or pair
annihilation.

We now consider a hole state (of momentum $\hbox{\boldmath$p$}$, $p<\mu$). 
The scattering process is now described in two
steps: first an initial vacancy in the Fermi sea is filled by a fermion of 
momentum
$\hbox{\boldmath$p'$}$; next the energy difference $p'-p$ is transferred to 
a fermion of momentum $\hbox{\boldmath$k'$}$, which is then extracted out of 
the Fermi sea. 
Hence, Fermi statistics imposes for this process the conditions
$0<k',p'<\mu<k$.
We define $\omega$ and $q$ as in eq.~(\ref{defwq}). 
Thus eqs.~(\ref{Gammaex11}) and (\ref{qlimit}) are still valid. 
However, eq.~(\ref{contrainteel}) is now replaced by 
\begin{equation}
\label{contrainteho}
p-\mu<\omega<\mu-k<0.
\end{equation}
With these conditions, eq.~(\ref{qlimit}) reduces to $-\omega<q<2p-\omega$. 
The phase space integration of eq.~(\ref{Gammaex11}) gives then 
\begin{eqnarray}
\Gamma(p)&=&\frac{3 e^4}{128 \pi^3}\ \frac{(\mu-p)^2}{p}. \label{Gamyu<}
\end{eqnarray}
Near the Fermi surface, the damping rate of fermions and holes 
vanishes quadratically with the excitation energy $|p-\mu|$. 
We shall come back to this in section~\ref{s:discussion}. 

Note that the damping rate diverges for small $p$. Extrapolating 
the above formula to the soft domain $p\sim e\mu$ (where our 
calculation does not apply), one guesses that 
$\Gamma$ is of order $e^3\mu$ for a soft excitation, instead of 
$e^4\mu$ for a hard excitation. A correct calculation shows that 
it is indeed the case \cite{VO}. 

\subsection{Antifermion in Yukawa's theory}
\label{s:antifermion yukawa}

Two collision processes contribute to leading order~: 
Bhabha scattering (Fig.~5~a) and pair annihilation (Fig.~5~b). 

The first one gives no difficulty. Its matrix element is
\begin{eqnarray}
\overline{|M|^2}&=&2 e^4,
\label{melbhayu}
\end{eqnarray}
and the phase space integration goes along the same lines as in 
section~\ref{s:fermion yuka}, 
except for the fact that no restriction applies on the final state energy $p'$.
Equation (\ref{contrainteel}) is therefore replaced by 
\begin{equation}
\label{contraintepo}
0<\mu-k<\omega.
\end{equation}
The contribution of Bhabha scattering
to the damping rate is then obtained 
by integrating eq.~(\ref{Gammaex11}) using eqs.~(\ref{qlimit}) and 
(\ref{contraintepo}):
\begin{eqnarray}
\Gamma_1(p)&=&\left\{\matrix{
\displaystyle\frac{e^4}{192 \pi^3} p\hfill & \hbox{\rm for} \ p < \mu, \cr
\displaystyle\frac{e^4}{192 \pi^3} \frac{\mu^2}{p^2} (3 p- 2 \mu) & 
\hbox{\rm for} \ 
p > \mu. }
\right.
\label{Gbhayu}
\end{eqnarray}

For pair annihilation, the tree matrix element is given by 
eq.~(\ref{melannyu}), and we need to include both hard and 
soft momentum transfers, 
according to the discussion following this equation. 
The hard sector contribution is given by eq.~(\ref{Gammaex11}). 
To use this equation, we must average the matrix element over 
$\phi$, the azimuthal angle of $\hbox{\boldmath$k$}$ with respect 
to $\hbox{\boldmath$q$}$. 
We first note that exchanging $t$ and $u$ in eq.~(\ref{melannyu}) 
amounts to exchanging the two outgoing photons, 
thus the $u/t$ and $t/u$ terms give identical 
contributions and $\overline{|M|^2}$ can be replaced by $e^4(u/t-1)$. 
   From the definition of $t$ and eq.~(\ref{defwq}), 
$t=Q^2=\omega^2-q^2$. The variable $u$ is given by 
$u=2(\hbox{\boldmath$p$}\cdot\hbox{\boldmath$k'$}-pk')$. 
Decomposing $\hbox{\boldmath$p$}$ and $\hbox{\boldmath$k'$}$ 
into longitudinal and transverse components with respect to 
$\hbox{\boldmath$q$}$, and averaging over $\phi$, one gets
$\left< \hbox{\boldmath$p$}\cdot\hbox{\boldmath$k'$}\right >=
(\hbox{\boldmath$p$}\cdot\hbox{\boldmath$q$})
(\hbox{\boldmath$k'$}\cdot\hbox{\boldmath$q$})/q^2$.
   From eq.~(\ref{defwq}), one obtains 
$\hbox{\boldmath$p$}\cdot\hbox{\boldmath$q$}=\omega p+(q^2-\omega^2)/2$
and $\hbox{\boldmath$k'$}\cdot\hbox{\boldmath$q$}=\omega k+(q^2+\omega^2)/2$. 
The average over $\phi$ gives therefore $\langle|M|^2\rangle=
e^4(\langle u/t \rangle -1)$, with 
\begin{equation}
\label{umoyen}
\left< {u \over t}\right> ={1\over 2 q^2}\left[(2 k+\omega)
(2p-\omega)-q^2\right].
\end{equation}
Notice that the collinear divergence (at $\omega \sim \pm q$) canceled out
of the ratio.

The limits on $q$ are given by eq.~(\ref{qlimit}), and the only 
additional restriction from Fermi statistics is $k < \mu$.
The integral is logarithmically divergent:
\begin{eqnarray}
\Gamma_{2h}(p)&=&\frac{e^4 \mu^2}{128 \pi^3 p}
\left(\log\frac{\mu p}{q_*^2}-\frac{3}{2}\right) ,
\label{Gannyuf}
\end{eqnarray}
where $q_*$ is an IR cutoff for the $q-$integral.

The soft sector contribution corresponds to
pair annihilation ``at low angles'', 
in the sense that the outgoing photons have momenta very close to those of the 
incoming electron and positron (Fig.~6, left). 
This can be viewed as a process where a hard antifermion ``turns'' into 
a hard boson (see Fig.~6, right) by absorbing a soft fermion ($\omega<0$) 
or emitting a soft antifermion ($\omega>0$).
The spectral function of a soft fermion receives contributions 
from two channels which are labeled by $+$ and $-$ in appendix A. 
We denote by $M_+$ and $M_-$ the corresponding matrix elements
\begin{equation}
\label{myukawa}
M_\pm=e\bar u (\hbox{\boldmath$p$},\lambda) u(\pm\hbox{\boldmath$q$},\lambda').
\end{equation}
Only states with opposite chiralities have non-vanishing matrix 
elements, which implies $\lambda'=-\lambda$. 
Using eq.~(\ref{spinormal}) and the property that 
$\left|\phi_{\hat{\hbox{\boldmath$p$}}}^\dagger
\phi_{\hat{\hbox{\boldmath$q$}}}\right| =\cos(\theta/2)$,
$\theta$ denoting the angle between 
$\hbox{\boldmath$p$}$ and $\hbox{\boldmath$q$}$, 
one easily obtains the result
\begin{equation}
\sum_{\lambda'}|M_{\pm}|^2=2e^2 p\, (q\mp\omega),
\label{Mpm2}
\end{equation}
where we have used the relation $\cos\theta=\omega/q$. 
  From eqs.~(\ref{gammasoftp}) and (\ref{Mpm2}), the soft contribution to 
the damping rate becomes 
\begin{eqnarray}
\label{gammasoft2}
\Gamma_s(p)={e^2\over 8\pi^2\, p}\int_{0}^{q_*} qdq\int_{-q}^{q}d\omega
\left[(q-\omega)\rho_+(\omega,q)+(q+\omega)\rho_-(\omega,q)\right],
\end{eqnarray}
or, by using the relationship $\rho_{+}(\omega,q)=\rho_-(-\omega,q)$, 
\begin{eqnarray}
\Gamma_{2s}(p)&=& \frac{e^2}{4 \pi^2 p}\int_0^{q_*}q dq\int_{-q}^{q}d\omega
\ (q-\omega)\ \rho_+(q,\omega).
\label{Gannsoftyu2}
\end{eqnarray}

To integrate over $\omega$, we first show the following sum rule:
\begin{equation}
\label{sumrule}
\int_{-\infty}^{\infty}d\omega\, (q-\omega)\rho_+(q,\omega)=0.
\end{equation}
Since $\rho_+$ is the discontinuity of the Green function $G_+$ on the 
real axis, the contour integral in eq.~(\ref{sumrule}) is simply the integral 
of $(q-z)G_+(q,z)$, with $z$ on a contour going parallel and right above the 
real axis from $-\infty$ to $+\infty$, and coming back right below it from 
$+\infty$ to $-\infty$.
Deforming the contour into a circle of infinite radius, 
on which $G_+(q,z)$ reduces to 
the free propagator $\left [2q(z-q)\right]^{-1}$, 
one easily sees that the integral vanishes.  
With the help of the sum rule of eq.~(\ref{sumrule}),
one can then express the contribution from the cut piece ($-q<\omega<q$)
of the density $\rho_{+}$ in eq.~(\ref{Gannsoftyu2}) in terms of the pole 
piece ($|\omega|>q$) of the density $\rho_{+}$. 
The pole contribution to the spectral density is $\rho_+=2\pi\delta(G_+^{-1})$.
Using eqs.~(\ref{Afreefermion}) and (\ref{ADysonfermion}), we get 
\begin{eqnarray}
\Gamma_{2s}(p)=
\frac{e^2}{4 \pi^2\, p}\int_0^{q_*}qdq \int_{|\omega|>q}d\omega
\,(\omega-q)
 2\pi\delta\Bigl(2q(\omega-q)-\Sigma_+(q,\omega)\Bigr).
\label{Gannsoftyu3}
\end{eqnarray}
Since the mass operator $\Sigma_+(q,\omega)$, given by eq.~(\ref{Asigma}), 
depends only on $\omega/q$, it is convenient to 
change variables from $q$, $\omega$ to $x=\omega/q$ and $y=2q(\omega-q)$, 
which gives
\begin{equation}
\Gamma_{2s}(p)=
\frac{e^2}{16 \pi\, p}\int {y dx dy\over |x-1|}
\delta\Bigl(y-\Sigma_+(x)\Bigr).
\label{Gannsoftyu4}
\end{equation}
This can be readily integrated over $y$ and then over $x$. 
The integration limits are derived from the dispersion relation 
eq.~(\ref{Adispfermion}). For $q\gg m_f$, one branch is at 
$x\simeq 1+m_f^2/q^2$
and the other is at $x\simeq -1$. Thus the integral on $x$ extends from 
$-\infty$ to $-1$ and from $1+m_f^2/q_{*}^2$ to $+\infty$. 
One gets 
\begin{eqnarray}
\Gamma_{2s}(p)&=&\frac{e^2 m_f^2}{8 \pi p}
\left[\log\left(\frac{2 q_*^2}{m_f^2}\right)-2\right]. 
\label{Gannsoftyuf}
\end{eqnarray}
Replacing $m_f$ by its value given in Appendix A 
and adding eqs.~(\ref{Gannyuf}) and (\ref{Gannsoftyuf}), the cutoff $q_*$
cancels out. The contribution of pair annihilation to the damping rate is thus
\begin{eqnarray}
\label{Gannyu}
\Gamma_2(p)&=&\Gamma_{2h}(p)+\Gamma_{2s}(p)\nonumber\\
&=&\frac{e^4 \mu^2}{128 \pi^3 p}\left[
\log\left( {2\mu p\over m_f^2}\right)-{7\over 2}\right]
\end{eqnarray}

The total damping rate to order $e^4$ is the sum of $\Gamma_1$ and 
$\Gamma_2$ given by Eqs.~(\ref{Gbhayu}) and (\ref{Gannyu}). 
It corresponds to the imaginary part of the self--energy diagrams displayed 
in Fig.~7. The two--loop diagrams give the hard contribution, 
while the (resummed) one--loop diagram gives the soft contribution. 

\subsection{Fermion and antifermion in QED and QCD}
\label{s:fermion antifermion}

We turn now to studying the fermion lifetime in relativistic QED and QCD 
plasmas. The damping process is fermion--fermion scattering and is 
dominated 
by soft momentum transfers, as explained in section~\ref{s:hard contribution}. 
In QED, we compute this process as 
 the emission of an off--shell soft photon by 
a hard electron (Fig.~2). 
The matrix element of the transition is 
$M=e\,\bar u(\hbox{\boldmath$p'$},\lambda')\gamma^\mu 
u(\hbox{\boldmath$p$},\lambda)\epsilon_\mu(\hbox{\boldmath$q$})$, 
where $\epsilon_\mu(\hbox{\boldmath$q$})$ is the polarization of the photon. 
Since $q\ll p$, one can replace 
$\hbox{\boldmath$p'$}=\hbox{\boldmath$p$}-\hbox{\boldmath$q$}$ by 
$\hbox{\boldmath$p$}$ in the 
matrix element. Then, the first factor in $M$ is simply the electric current 
associated with the incoming electron, which reduces to 
$J^\mu=e\,\bar u(\hbox{\boldmath$p$},\lambda')\gamma^\mu 
u(\hbox{\boldmath$p$},\lambda)=2e\, P^\mu\delta_{\lambda,\lambda'}$,
the Kronecker symbol reflecting the fact that helicity is conserved
in the process. We thus obtain $M=2 e\, P\cdot\epsilon$. 
Now, in the Coulomb gauge, the spectral function of the soft photon 
receives a contribution from longitudinal  and transverse modes, 
which are denoted by $\rho_L(q,\omega)$ and $\rho_T(q,\omega)$ 
respectively (see Appendix A). 
The longitudinal polarization vector gives a matrix element 
$\overline{|M_L|^2}=4 e^2 p^2$. 
The two transverse polarization vectors $\epsilon_T^\mu(q,\lambda)$ 
satisfy 
$\sum_{\lambda=1,2}\epsilon_T^i(q,\lambda)\,\epsilon_T^j(q,\lambda)=
\delta_{i j}-q_i q_j/q^2$ which gives, upon using $\omega=q \cos\theta$, 
$\overline{|M_T|^2}=4e^2 p^2(1-\omega^2/q^2)$. 
The soft contribution to the damping rate is given by eq.~(\ref{gammasoftp}) 
and the condition $0<\omega<p-\mu$:
\begin{eqnarray}
\label{gammasoft1}
\Gamma_s(p)={e^2\over 4\pi^2}\int_{0}^{q_*} qdq
\int_{0}^{\min(q,p-\mu)}d\omega
\left[\rho_L(\omega,q)+\left(1-{\omega^2\over q^2}\right)
\rho_T(\omega,q)\right].
\end{eqnarray}
The restriction $\omega>0$ is due to the conditions $k<\mu$ and 
$k'>\mu$, which together imply $\omega=k'-k>0$ (see Fig.2, left). 
The damping of a hole is calculated in the same manner and leads to the 
same expression eq.~(\ref{gammasoft1}), with $p-\mu$ replaced by $\mu-p$.

Since the spectral functions $\rho_L$ and $\rho_T$ fall off rapidly 
for $q>e \mu$, one can safely take the cutoff $q_*$ to infinity. 
Introducing then 
the dimensionless quantities $y=q/q_D$ in eq.~(\ref{gammasoft1}), 
$\varepsilon=|p-\mu|/q_D$, 
and $x=\omega/q$, one obtains
\begin{eqnarray}
\label{gammasoft11}
\Gamma(\varepsilon)&=& \frac{e^2 q_D}{4 \pi^2}\int_0^1 dx
\int_{0}^{\varepsilon/x} y^2 dy\left[r_L(y,x)+(1-x^2) r_T(y,x)\right],
\label{Gamfersoft21}
\end{eqnarray}
with $r_{L,T}=q_D^2\ \rho_{L,T}$. 

In QCD, the damping rate of a quark or a quark hole has a similar 
expression because the color current of a quark (which enters the 
matrix element) has the same structure as its electric current. 
The result is then given by eq.~(\ref{gammasoft11}) with a 
multiplicative color factor 
$C_f=(N_c^2-1)/(2N_c)$: $N_c^2-1$ for the number of soft gluons, 
$1/N_c$ for the average over the quark colors and $1/2$ for 
the trace of SU(3) generator products.

The integrals in eq.~(\ref{gammasoft11}) can be evaluated 
numerically. Simple approximate results can be derived in two limits:

1. Far from the Fermi surface, $\varepsilon \gg 1$. As the spectral densities
$r_L$ and $r_T$ vanish rapidly for $y\ge 1$, we can extend the upper bound 
of the $y-$integral to infinity. One finds
\begin{eqnarray}
\Gamma(p)&=&0.057 e^2\, C_f q_D,
\label{eq:gaminf}
\end{eqnarray}
or 
\begin{eqnarray}
\label{Gfermion}
\Gamma(p)&=&\left\{
\matrix{
\displaystyle 0.018 e^3 \mu & {\rm for\ QED,}\hfill\cr 
\displaystyle 0.017 \sqrt{N_f} e^3 \mu& {\rm for\ QCD\ with\ }
N_f{\rm\ flavors,}\ N_c=3.
}\right.
\end{eqnarray}

2. Very close to the Fermi surface, $\varepsilon \ll 1$. 
The integration over $y$ covers an appreciable range of values 
for small values of $x\sim\varepsilon$ only, so that one can safely
extend the integral over $x$ to infinity.
In the static limit, $x=\omega/q \ll 1$, 
the longitudinal and transverse spectral functions 
behave very differently.  
Static electric fields are screened at distances larger than $q_D^{-1}$.
The longitudinal 
polarization function, given by eq.~(\ref{PiL}), reduces to a constant, 
$\Pi_L\simeq -q_D^2$, and 
the corresponding spectral function $r_L$ is 
\begin{equation}\label{r_L}
r_L=\frac{\pi x}{(y^2+1)^2}.
\end{equation} 
Integrating over $x$ first, one finds a contribution 
to the damping rate of order $\varepsilon^2$
\begin{eqnarray}
\Gamma_L= 
C_f \frac{e^2 q_D}{4 \pi}\int_0^{\infty} dy \int_0^{\varepsilon/y} dx\,
 \frac{x y^2}{(y^2+1)^2}+{O}(\varepsilon^3) &\sim& 
C_f \frac{e^2 q_D}{32} \varepsilon^2 +{O}(\varepsilon^3).
\label{rl}
\end{eqnarray}
On the contrary, the transverse polarization function is purely imaginary for 
$x\ll 1$, because a static magnetic field is not screened. 
In this limit ($x=\omega/q \ll 1$), eq.~(\ref{PiT}) gives 
$\Pi_T=-i\pi q_D^2\omega/4q^3$, 
and the spectral function $r_T$ is 
\begin{equation}\label{r_T}
r_T=\frac{\pi x}{2 (y^4+\pi^2 (x/4)^2)}.
\end{equation}
Even though there is no static screening, the term proportional to $x^2$ 
in the denominator induces a deviation from the Rutherford 
$1/q^4$ ({\it i.e.} $1/y^4$) behavior, 
which is referred to as dynamical screening.  
The main contribution to the damping rate comes from values of $y$ and $x$ 
($x\sim\varepsilon$) such that $y^4$ and $x^2$ are of the same order, 
i.e. such that $q\sim (q_D^2\omega)^{1/3}$. 
Upon introducing the variable $y=y'\sqrt{x}/2$ and integrating over
$x$ first, one finds
\begin{eqnarray}
C_f \frac{e^2 q_D \varepsilon}{3 \pi}\int_0^{\infty} dy' 
\frac{y'}{y^{\prime 4}+\pi^2}
+{O}(\varepsilon^3)&\sim&
 C_f \frac{e^2 q_D}{12 \pi} \varepsilon+{O}(\varepsilon^3). 
\label{rt}
\end{eqnarray}
The damping rate is thus dominated by the transverse contribution and gives
\begin{eqnarray}
\label{GelQEDfs}
\Gamma(p)&=& \left\{
\matrix{
\displaystyle \frac{e^2}{12 \pi} |p-\mu|&{\rm for\ QED,}\hfill\cr
\displaystyle \frac{e^2}{9 \pi} |p-\mu|&{\rm for\ QCD\ with\ }N_f
{\rm\ flavors,}\ N_c=3.
}\right.
\end{eqnarray}

For the damping rate of a positron or an antiquark, Rutherford scattering has
a matrix element varying as
$(u/t)^2$ and dominates over pair--annihilation and the $s-$channel
contribution to Bhabha scattering, varying respectively as $(u/t)$ and 
$(u^2/s^2)$. 
The soft sector contribution gives
therefore the leading order term. Since the electric and color
currents are the same as for electrons or quarks, up to a sign, 
the damping rate is also given by eq.~(\ref{gammasoft11}), 
without the restriction $t<\epsilon/x$ coming from Pauli blocking. 
The damping rate is then given by the result of eq.~(\ref{Gfermion}) for 
all momenta.

\subsection{Boson in Yukawa's theory and QED}
\label{s:boson}

An incident beam (of given central energy) of scalar bosons or photons will 
undergo a spectral broadening due to the elastic scattering of its quanta
with electrons (Compton scattering). 
The matrix element squared, $\overline{|M|^2}$, corresponding to the diagrams 
depicted in Fig.~8, is related to that of pair annihilation 
by crossing symmetry. 
For a Yukawa interaction, one deduces from eq.~(\ref{melannyu}) 
\begin{eqnarray}
\overline{|M|^2}&=&2 e^4 \left( -\frac{u}{s}-\frac{s}{u}+2\right),
\label{melcomptonyu}
\end{eqnarray}
whereas for a hard photon we have
\begin{eqnarray}
\overline{|M|^2}&=&4 e^4 \left(-\frac{u}{s}-\frac{s}{u}\right).
\label{melcomptonQED}
\end{eqnarray}
The computation of the damping rate is similar to the calculation 
done in section~\ref{s:antifermion yukawa}. In particular, 
both hard and soft momentum transfers may contribute.

To compute the hard contribution, we define the 
variables $\omega=p+k=p'+k'$ and 
$\hbox{\boldmath$q$}=\hbox{\boldmath$p$}+\hbox{\boldmath$k$}
=\hbox{\boldmath$p'$}+\hbox{\boldmath$k'$}$, 
so that
$s=\omega^2-q^2$. Taking $k'$, $\omega$ and $q$ as integration variables, 
the contribution to the 
damping rate is given by eq.~(\ref{Gammaex11}), with $k$ replaced by 
$k'$. The matrix element $\overline{|M|^2}$ 
must be averaged over the azimuthal 
angle of $\hbox{\boldmath$k$}$ around $\hbox{\boldmath$q$}$. 
Following the same method as for pair 
annihilation, we obtain an equation similar to eq.~(\ref{umoyen}):
\begin{equation}
\left< {u\over s}\right> = {1\over 2q^2}
\left[ (2k'-\omega)(2p-\omega)-q^2\right].
\end{equation}
The limits on $q$ are $\max(|2p-\omega|,|2k'-\omega|)<q<\omega$, 
and the constraints from Fermi statistics are $\omega-p<\mu<k'$. 

The inverse term $(-s/u)$ is integrated using the variables 
$\omega=k-p'=k'-p$, 
$\hbox{\boldmath$q$}=\hbox{\boldmath$k$}-\hbox{\boldmath$k'$}
=\hbox{\boldmath$k'$}-\hbox{\boldmath$p$}$ and following 
the same steps as for the direct term. 

Adding up the direct and inverse terms, we find for the hard sector 
contribution to the damping rate of a Yukawa boson
\begin{equation}
\matrix{
\Gamma(p<\mu)=
\displaystyle\frac{e^4 p}{192 \pi^3}
\left[1-\frac{3 \mu^2}{p^2}\log\left(1-\frac{p^2}{\mu^2}\right)\right]
\hfill\cr
\Gamma(p>\mu)=
\displaystyle\frac{e^4 \mu^2}{64 \pi^3 p} 
\left[3-\frac{2\mu}{3 p}+\log\left(\frac{
(p-\mu) p \mu}{(p+\mu) q_*^2}\right)\right].\hfill
}
\label{Ggamyu}
\end{equation}
For a photon (QED) we obtain
\begin{equation}
\matrix{
\Gamma(p<\mu)=
\displaystyle -\frac{e^4 p}{96 \pi^3}\left[1+\frac{3 \mu^2}{p^2}
\log\left(1-\frac{p^2}{\mu^2}\right)\right]\hfill\cr
\Gamma(p>\mu)=
\displaystyle\frac{e^4 \mu^2}{32 \pi^3 p}
\left[1+\frac{2 \mu}{3 p}+\log\left(\frac{(p-\mu) p \mu}{(p+\mu) q_*^2}\right)
\right].\hfill}
\label{GgamQED}
\end{equation}
As in section~(\ref{s:antifermion yukawa}), the 
logarithmic divergence in $q_*$ corresponds to processes where 
the intermediate fermion state is soft in the second diagram of Fig.~8. 
In these processes, the incoming photon (boson) transfers 
almost all its energy to the outgoing electron. 
Since the outgoing electron is always above the Fermi level, 
this situation can occur only if $p>\mu$
(We assume for simplicity that $|p-\mu|\gg e \mu$). 
Thus the logarithmic divergence is present only for $p>\mu$, as can 
be seen in eqs.~(\ref{Ggamyu}) and (\ref{GgamQED}). 
Then, both soft and hard momentum transfers contribute. 

The soft sector contribution corresponds to the emission of a soft 
antifermion
(or to the absorption of a soft fermion) by a hard 
boson (Fig.~9, right). 
Since $k'\simeq p$, the corresponding matrix element is 
approximately given by eq.~(\ref{myukawa}) for a Yukawa boson, 
and by 
\begin{equation}
\label{mQED}
M_\pm=e\,\bar u (\hbox{\boldmath$p$},\lambda)
\, \hbox{\boldmath$\epsilon$}_T\cdot\hbox{\boldmath$\gamma$}
\, u(\pm\hbox{\boldmath$q$},\lambda'),
\end{equation}
for a QED photon, where 
$\hbox{\boldmath$\epsilon$}_T$ is the polarization vector. 
Since $\hbox{\boldmath$\epsilon$}_T$ is transverse with respect to 
the photon momentum $\hbox{\boldmath$p$}$, 
The operator $\hbox{\boldmath$\epsilon$}_T\cdot\hbox{\boldmath$\gamma$}$
anticommutes with the Dirac operator 
$p\,\gamma^0-\hbox{\boldmath$p$}\cdot\hbox{\boldmath$\gamma$}$ 
and with the chirality operator $\gamma^5$. 
Thus, it simply changes $\bar u(\hbox{\boldmath$p$},\lambda)$ into 
$\bar u(\hbox{\boldmath$p$},-\lambda)$, up to a phase. 
The matrix element is the same as for the Yukawa interaction.  
The only difference is that the chirality $\lambda$ is conserved 
($\lambda'=\lambda$), while it changes in the Yukawa theory 
($\lambda'=-\lambda$). 

In both cases, the resulting soft sector contribution
takes therefore the  same form as the soft contribution for positron 
annihilation in eq.~(\ref{Gannsoftyu2}) with an additional factor $2$ 
for the final electron spins, and with an additional restriction 
on phase space from Fermi statistics, $p-\omega>\mu$. 
Since $\omega$ is of order $ e\mu$, this restriction can be ignored as 
soon as the boson energy is not too close to the Fermi 
energy, i.e. if $p-\mu\gg e\mu$. 
In this condition, the damping rate is given by 
eq.~(\ref{Gannsoftyuf}), multiplied by a factor of 2. 
Adding up the hard and soft contributions, we find for the hard 
Yukawa boson   
\begin{equation}
\label{Ggamfinalyuka}
\matrix{
\Gamma(p<\mu)=
\displaystyle \frac{e^4 p}{192\pi^3 }\left[1-\frac{3\mu^2}{p^2}
\log\left(1-{p^2\over \mu^2}\right)\right]\hfill\cr
\Gamma(p>\mu)=
\displaystyle \frac{e^4 \mu^2}{64 \pi^3 p}\left[1-\frac{2 \mu}{3 p}+
\log\left(\frac{2 p \mu (p-\mu)}{m^2_f (p+\mu)}\right)\right],\hfill}
\end{equation}
and for a hard photon
\begin{equation}
\label{GgamfinalQED}
\matrix{
\Gamma(p<\mu)=
\displaystyle -\frac{e^4 p}{96 \pi^3}\left[1+{3\mu^2\over p^2}
\log\left(1-{p^2\over \mu^2}\right)\right]\hfill\cr
\Gamma(p>\mu)=
\displaystyle \frac{e^4 \mu^2}{32 \pi^3 p}\left[-1+\frac{2 \mu}{3 p}+
\log\left(\frac{2 p \mu (p-\mu)}{m^2_f (p+\mu)}\right)\right]. \hfill }
\end{equation}
Once again the cutoff $q_*$ has canceled out upon addition of the hard and 
soft contributions. 
These results are valid only far from the Fermi surface, i.e. 
for $|p-\mu|\gg e\mu$. 
The corresponding self--energy diagrams are displayed in Fig.~10. 

\subsection{Gluon}
\label{s:gluon}

Three tree diagrams contribute to Compton scattering of a gluon. They
are displayed in Fig.~11. The first, as shown below, yields a 
contribution of order 
$e^3\mu$; the contribution from the other ones is subleading and
is of the same order in $ e$ as in QED, {\it i.e.} 
of order $e^4\mu$ or $e^4\mu\log(1/ e)$ (see previous subsection). 

The scattering process in Fig.~11, left, can be viewed as the emission of 
a soft gluon by a hard one (see Fig.~12). The matrix element is evaluated as
follows.
The three gluon vertex, coupling a hard gluon of color 
index $a$, momentum $p$ and polarization $\epsilon_p$ to a hard and a soft 
gluon of color indices $b$ and $c$, momenta $p'$ and $q$ and polarizations 
$\epsilon_{p'}$ and 
$\epsilon_q$ respectively, is
\begin{eqnarray}
\label{3v}
M=ef^{abc}\left(\epsilon_{p}\cdot\epsilon_{p'}\ (P+P')\cdot\epsilon_q + 
\epsilon_{p'}\cdot \epsilon_q\ (-P'+Q)\cdot \epsilon_{p}
+\epsilon_q\cdot \epsilon_{p}\ (-Q-P)\cdot \epsilon_{p'}\right),
\end{eqnarray}
where $f^{abc}$ is the $SU(3)$ structure constant. 
The hard gluons are on--shell transverse gluons, 
whose polarization vectors satisfy 
$\epsilon_p\cdot P=\epsilon_{p'}\cdot P'=0$. 
Therefore, in the limit $q \ll p$, the last two terms 
in eq.~(\ref{3v}) vanish. 
The remaining term is 
\begin{eqnarray}
\label{1v}
M=2e f^{abc}\left( \epsilon_{p}\cdot \epsilon_{p'}\right)
\left( P\cdot \epsilon_q\right). 
\end{eqnarray}
This can be written in the form $M=J_\mu\epsilon_q^{\mu}$
where $J_\mu=2 e f^{abc} (\epsilon_{p}\cdot \epsilon_{p'})P_\mu$ 
is the matrix element of the color current between 
the initial and final hard gluon states. In this form, it is analogous 
to the matrix element obtained in the case of the 
emission of a soft photon by a hard electron (see 
section~\ref{s:fermion antifermion}). 

The gluon damping rate is therefore given by an equation similar to 
eq.~(\ref{gammasoft1}), with two minor modifications: 
the condition $\omega<p-\mu$ does not apply for a final gluon state, 
and the result must be 
multiplied by a factor $N_c/2=f_{abc} f_{abc}/16$ 
coming from the color degrees of freedom: 
$N_c$ for the possible ways of transferring color to the soft gluons, 
and 1/2 for the symmetry factor. 
For $N_c=3$, the 
numerical value is
\begin{eqnarray}
\label{dgluon}
\Gamma(p)&=& \frac{3}{2}\ 0.057\ e^2\ q_D, \nonumber \\
&=& 0.019 \sqrt{N_f}\ e^3 \mu.
\end{eqnarray}

\section{Discussion}
\label{s:discussion}

Damping rates of hard particles in a cold ultrarelativistic fermion gas are 
at least of order $e^3$ higher than their energy: 
hence, one-particle excitations are narrow quasiparticle states in 
the perturbative regime $ e\ll 1$. 
The damping rates of the various one--particle excitations are displayed 
as a function of their momentum $p$ in Figs.~14, 15 and 16 for the 
Yukawa interaction, QED and QCD respectively. 

We distinguish two categories of damping processes:
for charged particles in gauge theories, 
the scattering process is essentially forward ($\theta\sim e$)
and the resulting damping rate is of order $e^3\mu$;
for other particles, large angle scattering ($\theta\sim 1$) contributes at 
least as much as small angle scattering, and the damping rate is of order 
$e^4\mu$ or $e^4\mu\log(1/e)$. 
Notice that the results derived in sections~\ref{s:fermion antifermion}, 
\ref{s:boson} and \ref{s:gluon} are gauge invariant: all the matrix 
elements used in either the hard or the soft sector contributions are averaged 
over physical polarization states.

\subsection{Damping rates of charged particles in gauge theories}

Damping rates of charged particles are dominated by collisions with soft 
momentum transfer, for which medium effects must be taken into account:
scattering takes place through the 
coupling of the elementary particle current with 
coherent plasma oscillations of the charge and current densities. 
The underlying classical structure is clear: the transition rate 
depends on the 
hard particle only through the associated current. 
We are in a situation where the hard particle motion is only slightly 
perturbed by the soft one: the gauge field 
behaves essentially as a classical field which couples to the 
current of the hard particles.

The electron and quark damping rates are very similar, in the sense that 
they differ only by 
trivial color factors. On the other hand, the gluon and photon damping 
processes are essentially different: the dominant contribution to gluon 
damping involves the three--gluon vertex (see Fig.~11), and the 
damping rate $\Gamma$ is of order $e^3\mu$, as for electrons and quarks. 
Gluon damping is therefore specifically non-abelian. 
The photon damping rate is smaller in magnitude, 
of order $e^4\mu\log(1/e)$ or $e^4\mu$. 

It is interesting to note that the damping rates of charged particles 
are independent of the particle momentum. The only exception comes from
electrons or quarks very close to the Fermi surface, within an interval 
$e\mu$ from the Fermi level, where the damping rate decreases with the 
excitation energy $|p-\mu|$ (section~\ref{s:fermion antifermion}). 
Then, the longitudinal part of the interaction 
is screened at low momenta and leads to a width {\it quadratic} in $|p-\mu|$. 
The dominant term, shown in eq.~(\ref{GelQEDfs}), is the contribution from
the transverse piece of the interaction, which is not screened in the static
limit, and the width is {\it linear} in $|p-\mu|$. This
correlation between the range of interaction and the electron damping rate
close to the Fermi surface is already well-known in the non--relativistic
electron gas (\cite{Lutt} and \cite{Norton}). In particular, our result of 
eq.~(\ref{GelQEDfs}) agrees with the energy dependence of the
imaginary part of the electron self--energy derived in \cite{Norton}, if we
set the Fermi velocity $v_F=p_F/m$ to $v_F=1$.

Note that the calculation presented here is valid only for hard momentum 
excitations. The damping of soft, charged, excitations in gauge theories 
has very different properties: first, it is momentum dependent; 
second, the hard contribution is no longer negligible, but becomes 
of the same order of magnitude as the soft contribution, yielding 
a damping rate of order $e^3\mu\log(1/ e)$, instead 
of $e^3\mu$ for hard excitations~\cite{VO}.

\subsection{Damping rates of other particles}

Other particles include neutral particles (photons, Yukawa scalars) and
fermions with a Yukawa coupling, for which the interaction  
$\phi\bar\psi\psi$ is not related to any conserved  charge.  
For these particles, the hard contribution to the damping rate is 
at least of the same order of magnitude as the soft contribution. 
While the soft contribution involves medium effects, i.e. coherent effects, 
the hard contribution simply results from incoherent collisions: 
a quasiparticle excitation dies off by kicking electrons out of the 
Fermi sea randomly. 
The resulting damping rates are smaller in magnitude, of order $e^4\mu$ or 
$e^4\mu\log(1/e)$. 
They are strongly momentum dependent, as can be seen in Figs.~14 and 15.

The $\log(1/e)$ term comes from processes in which a massless 
fermion is exchanged, and the fermion propagator must be corrected 
for medium effects. These processes are specific to relativistic 
plasmas. As a result of the medium effects,
the photon and the scalar boson damping rates are strongly momentum
dependent: 
they rise steeply near the Fermi energy (see Figs. 14 and 15), 
and are of order $e^4\mu$ for $p<\mu$ and $e^4\mu\log(1/e)$ for $p>\mu$. 
Once again, the soft momentum transfer process is almost classical in 
nature. Here, it is the 
fermionic soft field which acts as a classical field (recall that 
the dispersion relation of soft fermions is the same for both 
interactions) in which the hard particles move \cite{BlaIan}. 

Near the Fermi surface, the  fermion damping rate in a Yukawa theory 
decreases {\it quadratically} with $|p-\mu|$, in contrast with the electron 
and quark damping rates. 
This is a consequence of the fact that hard momentum transfers dominate
and that in this sense, the interaction is short--ranged. 

Damping rates of soft excitations are of order $e^3\mu$ or 
$e^3 \mu \log(1/e)$, i.e.
one power in $e$ smaller than for hard momenta. The additional factor $1/e$
comes from kinematics~\cite{VO}.

\subsection{Comparison with the high temperature case}

As noted in the introduction, ultrarelativistic plasmas have 
the same screening properties in the high density ($T=0$) and high 
temperature ($\mu=0$) limits. 
For neutral particles, we have seen that damping rates are of order
$e^4\mu$ or $e^4\mu\log(1/e)$ at $T=0$. These damping rates 
are generally of order $e^4 T\log(1/e)$ at high temperature\cite{Thomaphoton}. 
The $\log(1/e)$ factor comes from processes in which a soft fermion
is exchanged, i.e. Compton scattering and pair annihilation at low 
angles. At $T=0$, these processes are not always possible 
(see sections~\ref{s:fermion yuka} and \ref{s:boson}) and the 
$\log(1/e)$ then disappears. 
Apart from this difference, the orders of magnitude of damping rates 
are the same in the hot and cold plasmas. 

For charged particles, the situation is very different. 
Naive perturbation theory yields an infrared divergent damping rate 
in the $T=0$ ($\mu > 0$) and $\mu=0$ ($T>0$) limits. 
However, the level of divergence is 
different in these two cases, because screening at the one--loop level
gives a finite damping rate, of order $e^3\mu$, if $T=0$, 
whereas in the high temperature case the damping rate is logarithmically
divergent and of order $e^2 T$.
In both cases, the damping processes are elastic collisions with the 
charges in the plasma. 
The essential difference lies in Pauli blocking: at zero temperature,
processes in which a boson with energy $\omega\sim e\mu$ 
is exchanged have a phase space proportional to $\omega$ (only electrons very
close to the Fermi surface participate in the collisions), 
whereas at high temperature, phase space is proportional to $T$.
This factor $\omega$ both contributes a factor $e$ and kills the 
divergence at $\omega=0$.

\section*{Acknowledgments}

After completion of this work, we learned that M. le Bellac and 
C. Manuel have recently calculated the damping rates of 
electrons and holes near the Fermi surface in QED and QCD \cite{LeBellac96}. 
We have benefited from discussions with a number of people. 
It is a pleasure to thank J.P. Blaizot 
and E. Iancu for useful comments on the manuscript as well as G. Baym for 
helpful remarks. This work has been supported in part by the US
National Science Foundation under Grant NSF PHY94-21309.

\appendix
\section{Spectral densities of soft modes}

In this appendix, we recall how boson and fermion propagators at low momenta
are modified by medium effects in an ultrarelativistic plasma. 

\subsection{Soft gauge field}

In the Coulomb gauge, rotational invariance allows one to 
decompose the photon propagator into a longitudinal (L) 
and a transverse (T) piece~\cite{Pisarski89a}:
\begin{eqnarray}
\label{Adecboson}
D^{\mu\nu}(\hbox{\boldmath$q$},z)&=&
D_L(q,z)\epsilon^\mu_L\epsilon^\nu_L 
+ D_T(q,z)\sum_{\lambda=1,2}
\epsilon^\mu_T(\hbox{\boldmath$q$},\lambda)
\epsilon^\nu_T(\hbox{\boldmath$q$},\lambda)
\end{eqnarray}
where $\epsilon^{\mu}_L=\delta^\mu_0$ and 
$\epsilon^{\mu}_T(\hbox{\boldmath$q$},\lambda)$, $\lambda=1,2$, 
are spacelike unit 
vectors mutually orthogonal and transverse to $\hbox{\boldmath$q$}$, 
therefore satisfying 
$\sum_{\lambda=1,2}\epsilon^i_T(\hbox{\boldmath$q$},\lambda)
\epsilon^j_T(\hbox{\boldmath$q$},\lambda)
=\delta_{ij}-q_iq_j/\hbox{\boldmath$q$}^2$.  
For the bare QED interaction, 
the decomposition of the propagator $D^{\mu\nu}_0(q,z)$
according to eq.~(\ref{Adecboson}) gives 
\begin{eqnarray}
\label{Afreeboson}
\matrix{
D_{0\,L}^{-1}(q,z)=q^2;& D_{0\,T}^{-1}(q,z)=z^2-q^2.}
\end{eqnarray}
For soft momenta $q\sim  e\mu$, the propagator is modified by medium effects:
\begin{equation}
\label{ADysonboson}
D^{-1}_{\mu\nu}(q,z)=D^{-1}_{0,\mu\nu}(q,z)-\Pi_{\mu\nu}(q,z).
\end{equation}
To leading order in $e^2$, the polarization tensor 
$\Pi_{\mu\nu}(q,z)$ is given by bubble diagrams corresponding to the 
photon coupling to electron--hole intermediate 
states. 
Decomposing $\Pi_{\mu\nu}$ according to eq.~(\ref{Adecboson}), 
one gets ($\eta \to +0$)~\cite{Silin,Weldon82}:
\begin{eqnarray}
\label{PiL}
{1\over q_D^2}\,\Pi_L(q,\omega+i\eta)&=&
-1+\frac{\omega}{2 q}\left[\log\left|\frac{\omega+q}{\omega-q}\right|
-i\pi\theta(q^2-\omega^2)\right], \\
\label{PiT}
{1\over q_D^2}\,\Pi_T(q,\omega+i\eta)&=&
\frac{\omega^2}{2 q^2}+
\frac{\omega (q^2-\omega^2)}{4 q^3}
\left[\log\left|\frac{\omega+q}{\omega-q}\right|
-i\pi\theta(q^2-\omega^2)\right] . 
\end{eqnarray} 
Notice that the full one--loop self--energy contains also diagrams 
corresponding to electron--positron intermediate states, however their 
self--energies 
are a power of $e$ smaller than those of eqs.~(\ref{PiL}), (\ref{PiT}), 
see~\cite{Silin,Weldon82}.
In eqs.~(\ref{PiL}), (\ref{PiT}), $q_D$ is the Debye screening momentum, 
given by the first entry of the following table.
The resummed propagator obtained from eqs.~(\ref{ADysonboson}--\ref{PiT}) 
is drawn as a photon propagator with a ``blob'' 
(see for instance Fig.~2).

The gluon propagator and gluon self--energy are diagonal in color 
indices, which we omit for brevity. In the strict Coulomb gauge, in the sense 
defined in the second reference of~\cite{BraPis}, 
the decomposition~(\ref{Adecboson}) holds
for the gluon propagator, and eqs.~(\ref{Adecboson}) through (\ref{PiT})
remain valid. The Debye screening momentum $q_D$ is given by the second entry
of the following table.
$$\vbox{
\offinterlineskip
\halign{
\strut \vrule \hfill \quad # \quad \hfill & \vrule \hfill \quad
$#$ \quad \hfill \vrule \cr
\noalign{\hrule}
\bf Theory & q_D \cr
\noalign{\hrule}
QED & e \mu / \pi \cr
\noalign{\hrule}
QCD & e \mu \sqrt{N_f}/(\pi\sqrt 2) \cr
\noalign{\hrule} 
}
}$$

The spectral density 
$\rho^{\mu\nu}(\hbox{\boldmath$q$},\omega)
=-2\,{\rm Im}\,D^{\mu\nu}(\hbox{\boldmath$q$},\omega+i\eta)$, ($\eta \to +0$)
can be decomposed, like the propagator in eq.~(\ref{Adecboson}), 
into longitudinal and transverse pieces, 
$\rho_{L,T}(q,\omega)=-2\,{\rm Im}\,D_{L,T}(q,\omega+i\eta)$. 
Their expressions are easily obtained from eqs.~(\ref{Afreeboson}--\ref{PiT}). 
For free fields, they reduce to $\rho_{0\,L}(q,\omega)=0$, 
$\rho_{0\,T}(q,\omega)=2\pi\delta(\omega^2-q^2)
(\theta(\omega)-\theta(-\omega))$: the 
only peaks are at $\omega=\pm q$ and correspond to transverse photons 
(gluons).  
For soft momenta, $\rho_L$ and $\rho_T$ are modified by medium effects. 
The peaks of $\rho_T$ are shifted towards higher values of $|\omega|$, and 
a peak appears in $\rho_L$ at the values of $\omega$ given by 
$D_{0,L}^{-1}(q,\omega)-\Pi_L(q,\omega)=0$, corresponding to plasmon modes. 
Note that the real parts of $D^{-1}_{L,T}(q,\omega)$ are 
even in $\omega$ for fixed $q$, so that the peaks always 
appear in pairs of opposite sign $\pm\omega$. 
In addition to these peaks, the spectral densities  
have a continuous part for $|\omega|<q$ coming 
from the imaginary part of the polarization in eqs.~(\ref{PiL}--\ref{PiT}), 
which corresponds physically to Landau damping: waves with $|\omega|<q$ 
lose their energy by accelerating fermions. 

\subsection{Soft fermion}

The fermion propagator can be decomposed on a basis of spinors 
in the following way (once again, we omit trivial color indices
for the quark propagator):
\begin{eqnarray}
\label{Adecfermion}
G(\hbox{\boldmath$p$},z)=
G_+(p,z)\sum_{\lambda=-1,1} 
u(\hbox{\boldmath$p$},\lambda) \bar u(\hbox{\boldmath$p$},\lambda)
+ G_-(p,z)\sum_{\lambda=-1,1} 
u(-\hbox{\boldmath$p$},\lambda) \bar u(-\hbox{\boldmath$p$},\lambda) . 
\end{eqnarray}
In this decomposition, $u(\hbox{\boldmath$p$},\lambda)$ denotes 
a solution of the free massless Dirac equation 
$(p\gamma^0-\hbox{\boldmath$p$}\cdot\hbox{\boldmath$\gamma$})
u(\hbox{\boldmath$p$},\lambda)=0$ 
with chirality $\lambda$, normalized according to the relation 
$\sum_{\lambda=-1,1}
u(\hbox{\boldmath$p$},\lambda)\bar u(\hbox{\boldmath$p$},\lambda)=
p\gamma^0-\hbox{\boldmath$p$}\cdot\hbox{\boldmath$\gamma$}$. 
Note that $u(\hbox{\boldmath$p$},\lambda)$ 
is a positive energy solution, while 
$u(-\hbox{\boldmath$p$},\lambda)$ 
is the corresponding negative energy solution 
with the same momentum $\hbox{\boldmath$p$}$: it corresponds to a positron 
(or an antiquark) with 
momentum $-\hbox{\boldmath$p$}$. 
An explicit expression of $u(\hbox{\boldmath$p$},\lambda)$ 
is most easily obtained 
in the chiral representation of Dirac matrices:
\begin{equation}
\label{spinormal}
u(\hbox{\boldmath$p$},+1)=\sqrt{2p}
\left( \matrix{\phi_{\hat{\hbox{\boldmath$p$}}}\cr 0} \right) \ \ 
u(\hbox{\boldmath$p$},-1)=\sqrt{2p}
\left( \matrix{0\cr\phi_{-\hat{\hbox{\boldmath$p$}}}} \right)
\end{equation}
where $\hat{\hbox{\boldmath$p$}}\equiv\hbox{\boldmath$p$}/p$ and 
$\phi_{\hat{\hbox{\boldmath$p$}}}$ 
is a two component spinor pointing in the direction 
of $\hat{\hbox{\boldmath$p$}}$, i.e. satisfying 
$\hbox{\boldmath$\sigma$}\cdot\hat{\hbox{\boldmath$p$}}
\phi_{\hat{\hbox{\boldmath$p$}}}
=\phi_{\hat{\hbox{\boldmath$p$}}}$, normalized to unity 
$\phi_{\hat{\hbox{\boldmath$p$}}}^\dagger\phi_{\hat{\hbox{\boldmath$p$}}}=1$. 
Note that 
$\phi_{\hat{\hbox{\boldmath$p$}}}\phi_{\hat{\hbox{\boldmath$p$}}}^\dagger=
(1+\hbox{\boldmath$\sigma$}\cdot\hat{\hbox{\boldmath$p$}})/2$ 

For the free Dirac propagator 
$G^{-1}_0(\hbox{\boldmath$p$},z)
=z\gamma^0-\hbox{\boldmath$p$}\cdot\hbox{\boldmath$\gamma$}$, 
the decomposition (\ref{Adecfermion}) gives 
\begin{equation}
\label{Afreefermion}
G^{-1}_{0\,\pm}(p,z)=2p(z\mp p).
\end{equation}
As expected, 
the poles of $G_+$ and $G_-$ are respectively the positive 
and negative energy solution of the free Dirac equation. 
For soft momenta $p\sim e\mu$, the fermion propagator is corrected 
by medium effects:
\begin{equation}
\label{ADysonfermion}
G^{-1}(p,z)=G_0^{-1}(p,z)-\Sigma(p,z). 
\end{equation}
The mass operator $\Sigma(p,z)$ can be decomposed according to 
eq.~(\ref{Adecfermion}), so that 
$G_{\pm}^{-1}=G_{0\,\pm}^{-1}-\Sigma_{\pm}$. 
To leading order in $e^2$, the components $\Sigma_{\pm}$ are given 
by ($\eta \to +0$)~\cite{Klimov81,Weldon82bis}  
\begin{eqnarray}
\label{Asigma}
{1\over 2 m_f^2}\Sigma_+(p,\omega+i\eta) &=&1 -{\omega-p\over 2 p}
\left[\log\left|{\omega+p\over\omega-p}\right|-i\pi\theta(p^2-\omega^2)\right],
\cr
\Sigma_-(p,\omega+i\eta) 
&=&-{\rm Re}\,\Sigma_+(p,-\omega+i\eta)+i\,{\rm Im}\,\Sigma_+(p,-\omega+i\eta).
\end{eqnarray}
The self--energy functions $\Sigma_{\pm}$ only include the coupling of 
the soft external fermion to intermediate fermion states through the 
absorption or the emission of a boson. These terms are the dominant ones for 
a soft external fermion~\cite{Klimov81,Weldon82bis}.
The quantity $m_f$ in eq.~(\ref{Asigma})
is the quasiparticle rest energy and is given in the 
following table:
$$\vbox{
\offinterlineskip
\halign{
\strut \vrule \hfil \quad # \quad \hfill & \vrule \hfil \quad $#$
\quad \hfil \vrule \cr
\noalign{\hrule}
\bf Theory & m_f \cr
\noalign{\hrule}
Yukawa &  e \mu/(4 \pi)\cr
\noalign{\hrule}
QED &  e \mu/(\pi\sqrt 8)\cr
\noalign{\hrule}
QCD &  e \mu /(\pi\sqrt 6)\cr
\noalign{\hrule}
}
}$$

The spectral density 
$\rho_F(\hbox{\boldmath$p$},\omega)
=-2\,{\rm Im}\,G(\hbox{\boldmath$p$},\omega+i\eta)$
can be decomposed like the propagator in eq.~(\ref{Adecfermion}), 
with $G_\pm(p,z)$ replaced by the corresponding spectral 
density $\rho_\pm(p,\omega)=-2\,{\rm Im}\,G_\pm(p,\omega+i\eta)$. 
For a free Dirac field, the spectral density reduces to 
$\rho_\pm(p,\omega)=2\pi\delta(\omega^2-p^2)\theta(\pm\omega)$. 
Medium effects modify the spectral density for soft momenta. 
The position of the peaks of $\rho_+$ are the solutions of 
$G_+^{-1}(p,\omega)=0$. Using eqs.~(\ref{Afreefermion}--\ref{Asigma}), 
one obtains the dispersion relation in terms of the parameter 
$x\equiv\omega/p$:
\begin{eqnarray}
\label{Adispfermion}
{p^2\over m_f^2}&=&{1\over x-1}-{1\over 2}\log\left({x+1\over x-1}\right),\cr
\omega &=& p x.
\end{eqnarray}
There are two peaks for a given $p$. One with $x>1$, which corresponds 
to the bare fermion state slightly shifted by its interaction with the
medium. 
Furthermore, a second peak appears for $x<-1$.  
It corresponds to a new fermionic excitation called ``plasmino''
which has no counterpart in non--relativistic plasmas. 
Finally, $\rho_+$ has a continuous part in the region $|\omega|<p$, 
which corresponds to the fermionic analogue of Landau damping. 
The density $\rho_-$ has the same properties, 
with $\omega$ replaced by $-\omega$. 

\section{Self--energy diagrams and cutting rules}

We show explicitly that the kinetic theory approach used in this 
paper is equivalent to the approach using the formalism of field theory. 
In field theory, the damping rate $\Gamma(p)$ is 
defined from the imaginary part of the self energy. In the case of 
a fermion, using the notations of Appendix~A, this relation 
reads ($\eta \to +0$)~\cite{Weldoncoupure} 
\begin{equation}
\Gamma(p)=-2\,{\rm tr}\left[ 
{\setbox0=\hbox{$p$}\dimen0=\wd0\setbox1=\hbox{/} 
\dimen1=\wd1\ifdim\dimen0>\dimen1\rlap{\hbox to \dimen0{\hfil/\hfil}}   
p\else\rlap{\hbox to \dimen1{\hfil$p$\hfil}}/\fi}
\,{\rm Im}\Sigma(\hbox{\boldmath$p$} ,p+i\eta)\right]/4p 
=-2\, {\rm Im}\,\Sigma_+(\hbox{\boldmath$p$},p+i\eta).
\end{equation}
Cutting rules (see for example~\cite{KS}) allow to express the 
imaginary part of a generic self--energy diagram as the rate of a 
scattering process, thus providing the equivalence with the kinetic 
theory approach used in this paper. 
We show in this appendix that the hard contribution to the damping 
rate corresponds to a two--loop self energy diagram where the loop 
momenta are hard (see Fig.~1) while the soft contribution corresponds 
to a one--loop self--energy diagram with a soft internal momentum (Fig.~2). 
We take the example of electron--electron scattering in QED. 
Our arguments can be easily extended to other scattering processes. 

We proceed as follows: 
we first show that the imaginary part of the one--loop resummed diagram 
depicted in Fig.~17 corresponds to the probability to emit a soft 
photon (second equality in Fig.~2). 
Then we show the equality displayed in Fig.~1, i.e. that the rate of 
electron--electron scattering corresponds to the imaginary part of 
a two--loop self--energy diagram. Note, however, that the interference 
term between the two Feynman diagrams of electron--electron scattering 
(Fig.~3, right) is not included here. It corresponds to the imaginary part 
of another two--loop diagram (Fig.~4, right). 
Finally, we show that the imaginary part of the one--loop resummed 
diagram corresponds to electron--electron scattering with a 
resummed interaction (first equality in Fig.~2). 

\subsection{One soft loop}

The contribution of the diagram in Fig.~17 to the fermion self--energy 
is given by 
\begin{equation}
\label{Boneloop}
\Sigma(\hbox{\boldmath$p$} ,z_p)=\int{d^3q\over (2\pi)^3}
\int_{-i\infty}^{+i\infty}{dz_q\over 2i\pi}
(-i e\gamma^\mu) G_0(\hbox{\boldmath$p$} -\hbox{\boldmath$q$},z_p-z_q)
(-i e\gamma^\nu) D_{\mu\nu}(\hbox{\boldmath$q$},z_q).
\end{equation}
In this expression, $z_p=\mu+i x$ with $x$ real,
$G_0$ is the fermion propagator, which coincides with the 
free propagator (\ref{Afreefermion}) for a hard fermion, 
and $D$ is the soft photon propagator given by 
eqs.~(\ref{ADysonboson}--\ref{PiT}). 
We write the internal propagators using the spectral representations:
\begin{eqnarray}
D_{\mu\nu}(\hbox{\boldmath$q$},z_q)&=&\int_{-\infty}^{+\infty}
{d\omega_q\over 2\pi}{\rho_{\mu\nu}
(\hbox{\boldmath$q$},\omega_q)\over z_q-\omega_q},\cr
G_0(\hbox{\boldmath$p'$},z_{p'})&=&\int_{-\infty}^{+\infty}
{d\omega_{p'}\over 2\pi}
{\rho_F(\hbox{\boldmath$p'$},\omega_{p'})\over z_{p'}-\omega_{p'}}, 
\end{eqnarray}
with $z_q=i x$, $z_{p'}=\mu+i x$ and 
$\hbox{\boldmath$p'$}=\hbox{\boldmath$p$}-\hbox{\boldmath$q$}$. 
A straightforward contour integration gives for the integral over $z_q$:
\begin{equation}
\label{Bzq}
\int_{-i\infty}^{+i\infty}{dz_q\over 2i\pi} {1\over z_q-\omega_q}
{1\over z_p-z_q-\omega_{p'}}=
-{1+n(\omega_q)-f(\omega_{p'})\over z_p-\omega_q-\omega_{p'}},
\end{equation}
where we have introduced the Bose--Einstein and Fermi--Dirac distribution 
functions $n(\omega)$ and $f(\omega)$ which, in the limit $T=0$, reduce to 
\begin{eqnarray}
\label{BEFD}
n(\omega)&=&{1\over e^{\omega/T}-1}=\theta(\omega)-1,\cr
f(\omega)&=&{1\over e^{(\omega-\mu)/T}+1}=\theta(\mu-\omega).
\end{eqnarray}
After analytic continuation of $z_p$ to $\omega_p+i\eta$,
the imaginary part of eq.~(\ref{Bzq}) becomes
\begin{equation}
\label{Bphase1}
\pi\delta(\omega_p-\omega_q-\omega_{p'})\left(1+n(\omega_q)-f(\omega_{p'})
\right).
\end{equation}
The imaginary part of the self--energy can thus be obtained from 
eq.~(\ref{Boneloop}) through replacing 
the internal propagators by their spectral functions and the Matsubara 
frequencies $z$ by real frequencies $\omega$, and multiplying by 
the occupation factor from eq.~(\ref{Bphase1}): 
\begin{eqnarray}
\label{Bimoneloop}
-2\, {\rm Im}\, \Sigma(\hbox{\boldmath$p$},p+i\eta)=
\int{d^4 Q\over (2\pi)^4}
(e\gamma^\mu)\rho_F(P-Q)( e\gamma^\nu)
\rho_{\mu\nu}(Q) \nonumber\\
\times\left(1+n(\omega_q)-f(\omega_p-\omega_q)\right), 
\end{eqnarray}
where we have introduced the four vectors 
$P=(p,\hbox{\boldmath$p$})$ and  
$Q=(\omega_q,\hbox{\boldmath$q$})$, and $\omega_p=p$.

Note that the occupation factors  can be rewritten as 
\begin{equation}
\label{Brelation1}
1+n(\omega_q)-f(\omega_{p'})=\left(1+n(\omega_q)\right)
\left(1-f(\omega_{p'})\right) +n(\omega_q)f(\omega_{p'}).
\end{equation}
The two terms correspond to the amplitudes for the 
direct and inverse process in Fig.~2 (right). 
If $\omega_p>\mu$ (particle excitation), only the direct process
contributes, when $0<\omega_q<\omega_p-\mu$.
If $\omega_p<\mu$ (hole excitation), on the other hand, 
only the inverse process contributes, when $\mu-\omega_p<\omega_q<0$.
Note that in this last case, the occupation factor is $nf=-1$; 
however, the boson spectral function also has an opposite sign 
for $\omega_q<0$ (see eqs.~(\ref{PiL}-\ref{PiT})) so that the 
global sign is unchanged.

Decomposing the spectral functions according to eqs.~(\ref{Adecboson}) 
and (\ref{Adecfermion}), and taking into account that 
$\rho_+(p,\omega)=2\pi\delta(\omega^2-p^2)\theta(\omega)$
for a hard fermion (the spectral function is the same as in the vacuum, 
and $\rho_-$ does not contribute since $\omega_p-\omega_q>0$), 
one obtains from eq.~(\ref{Bimoneloop}):
\begin{eqnarray}
\label{Bimdec}
-2\, \bar u(\hbox{\boldmath$p$},\lambda)\,
{\rm Im}\, \Sigma(\hbox{\boldmath$p$},p+i\eta)u(\hbox{\boldmath$p$},\lambda)=
\int{d^4Q\over (2\pi)^4}
2\pi\delta((P-Q)^2)\sum_{j=L,T}\rho_j(Q) 
\sum_{\lambda'} \left| M_j\right|^2\nonumber\\ 
\times\left(1+n(\omega_q)-f(\omega_p-\omega_q)\right).
\end{eqnarray}
where $M_j$ is the matrix element of the transition process in Fig.~2 (right): 
\begin{equation}
M_j=\bar u(\hbox{\boldmath$p'$},\lambda') (-i e\gamma_\mu) 
u(\hbox{\boldmath$p$},\lambda)\epsilon^\mu_j(\hbox{\boldmath$q$}) .
\end{equation}
Comparing eq.~(\ref{Bimdec}) with eq.~(\ref{gammasoft}), one concludes
\begin{eqnarray}
\Gamma(p)&=&-2\, \bar u(\hbox{\boldmath$p$} ,\lambda)
{\rm Im}\, \Sigma(\hbox{\boldmath$p$} ,p+i\eta) 
u(\hbox{\boldmath$p$} ,\lambda) /2p,\\
&=&-2\,{\rm tr}\left[ 
{\setbox0=\hbox{$p$}\dimen0=\wd0\setbox1=\hbox{/} 
\dimen1=\wd1\ifdim\dimen0>\dimen1\rlap{\hbox to \dimen0{\hfil/\hfil}}   
p\else\rlap{\hbox to \dimen1{\hfil$p$\hfil}}/\fi}
\, {\rm Im}\Sigma(\hbox{\boldmath$p$} ,p+i\eta)\right]/4p. 
\end{eqnarray} 

\subsection{Two hard loops}

We now turn to the second step: we show that the imaginary part of 
the two--loop diagram on the right of Fig.~1 corresponds to the 
rate of electron--electron scattering with the diagram on the 
left of Fig.~1. 
The contribution of the two--loop diagram to the self--energy is 
given by a formula analogous to eq.~(\ref{Boneloop}), where 
$D$ is replaced by a photon line with a fermion loop insertion, 
i.e. by $D_0\Pi D_0$, with 
\begin{equation}
\label{Bpimunu}
\Pi^{\mu\nu}(\hbox{\boldmath$q$},z_q)=\int{d^3k\over (2\pi)^3}
\int_{\mu-i\infty}^{\mu+i\infty}{dz_k\over 2i\pi}
{\rm tr}\left[G_0(\hbox{\boldmath$k$},z_k)(e\gamma^\mu) 
G_0(\hbox{\boldmath$k$}+\hbox{\boldmath$q$},z_k+z_q)
(e\gamma^\nu)\right] 
\end{equation}
with $z_q$ on the imaginary axis. 
Therefore, the imaginary part of the two--loop diagram is given by 
eq.~(\ref{Bimoneloop}), in which the boson spectral function 
$\rho(\hbox{\boldmath$q$},\omega_q)
=-2\,{\rm Im}\,D(\hbox{\boldmath$q$},\omega_q+i\eta)$ 
is replaced by $-2\,{\rm Im}(D_0\Pi D_0)$. 
Now, the imaginary part of $D_0$ vanishes because both fermions 
are on mass shell ($|\omega_p|=p$ and 
$|\omega_p-\omega_q|=|\hbox{\boldmath$p$}-\hbox{\boldmath$q$}|$), 
which implies $|\omega|<q$. 
Thus we can write ${\rm Im}(D_0\Pi D_0)=\,D_0({\rm Im}\Pi)D_0$. 
To calculate ${\rm Im}\Pi$, 
we follow the same steps as for ${\rm Im}\Sigma$ in eq.~(\ref{Boneloop}). 
Using the spectral representation to write the internal propagators, 
the integral over $z_k$ can be calculated easily 
($\hbox{\boldmath$k'$} = \hbox{\boldmath$k$} +
\hbox{\boldmath$q$}$):
\begin{equation}
\label{Bzk}
\int_{\mu-i\infty}^{\mu+i\infty}{dz_k\over 2i\pi} {1\over z_k-\omega_k}
{1\over z_k+z_q-\omega_{k'}}=
{f(\omega_k)-f(\omega_{k'})\over z_q+\omega_k-\omega_{k'}}. 
\end{equation}
After analytic continuation of $z_q$ to $\omega_q+i\eta$, the imaginary 
part of this equation becomes
\begin{equation}
\label{Bphase2}
-\pi\delta(\omega_q+\omega_k-\omega_{k'})\left(f(\omega_k)-f(\omega_{k'})
\right).
\end{equation}
One thus obtains 
\begin{eqnarray}
\label{Bimpi}
-2\,{\rm Im}\,\Pi^{\mu\nu}(\hbox{\boldmath$q$},\omega_q+i\eta)
&=&\int{d^3k\over (2\pi)^3}
{d\omega_k\over 2\pi}
{\rm tr}\left[\rho_F(\hbox{\boldmath$k$},\omega_k)(e\gamma^\mu)
\rho_F(\hbox{\boldmath$k$}+\hbox{\boldmath$q$},
\omega_k+\omega_q)(e\gamma^\nu)\right]\nonumber\\
& & \times (f(\omega_k)-f(\omega_{k'})). 
\end{eqnarray}
Replacing $\rho$ in eq.~(\ref{Bimoneloop}) by 
$-2 D_0({\rm Im}\Pi)D_0$, using eq.~(\ref{Bimpi}), and decomposing the 
fermion spectral functions according to eq.~(\ref{Adecfermion}), 
one obtains
\begin{eqnarray}
\Gamma(p)&=&{1 \over {2 p}}\int{d^4Q\over (2\pi)^4}\int{d^4K\over (2\pi)^4}
\,(2\pi)\delta((P-Q)^2)\,(2\pi)\delta(K^2)\,(2\pi)\delta((K+Q)^2) 
\left|M\right|^2\nonumber\\
& & \times
\left(f(\omega_k)-f(\omega_{k'}))(1+n(\omega_q)-f(\omega_{p'})\right).
\end{eqnarray} 
where $K=(\omega_k,\hbox{\boldmath$k$})$. Notice that, from $K^2=0$,
the only possible solution for $\omega_k$ is $\omega_k=k$, only electrons are
initially present in the Fermi sea. The matrix element M is given by
$M=\sum_{\lambda',\kappa,\kappa'}J_{p',\lambda';p,\lambda}^{\mu}
J_{k',\kappa';k,\kappa\,\mu}(1/Q^2)$ with
$J_{p',\lambda';p,\lambda}^{\mu}=e \overline{u}(p',\lambda')\gamma^{\mu}
u(p,\lambda)$ and corresponds indeed to the direct contribution to
electron--electron scattering (Fig.~1, left). Introducing the four vectors
$K'=K+Q$ and $P'=P-Q$, integrating over $Q$ and making use of
the identity $\int (d^4 K)/(2\pi)^4 2\pi\delta(K^2)\Theta(K_0)=\int d\tau_k$, one finds 
eq.~(\ref{Gammaex1}), up to the statistical factors.
Thus, we only need to check that the product of the 
phase space factors eq.~(\ref{Bphase1}) and (\ref{Bphase2}) 
corresponds to electron--electron scattering. 
For this purpose, we note that the energy conservation 
$\omega_q+\omega_k=\omega_{k'}$ implies the following relations 
between the statistical factors:
\begin{eqnarray}
\label{einstein}
n(\omega_q)\left( f(\omega_k)-f(\omega_{k'}) \right) &=&
\left( 1-f(\omega_k)\right)f(\omega_{k'})\cr
\left(1+n(\omega_q)\right)\left( f(\omega_k)-f(\omega_{k'}) \right) &=&
\left( 1-f(\omega_{k'})\right)f(\omega_k)
\end{eqnarray}
Using these equations together with eq.~(\ref{Brelation1}), one obtains 
immediately the phase space factor under the form 
\begin{equation}
\label{stattrick}
(1-f(\omega_{p'}))f(\omega_k)(1-f(\omega_{k'}))
+f(\omega_{p'})(1-f(\omega_k))f(\omega_{k'}).
\end{equation}
The two terms correspond to the amplitudes of the direct and inverse 
process, as expected. 

\subsection{Screened interaction}

Finally, we show that the imaginary part of the one loop resummed 
diagram corresponds to the probability of electron--electron scattering 
with a screened interaction (first equality in Fig.~2). 
This is a straightforward extension of the previous result. 
We start from eq.~(\ref{Bimoneloop}). 
The spectral function of the resummed photon line, 
$\rho$, is given by 
$\rho=-2 {\rm Im}D=2 |D|^{2}({\rm Im}D^{-1})$. 
Using eq.~(\ref{ADysonboson}) and the fact that ${\rm Im}D_0^{-1}=0$, 
as discussed above, one obtains 
$\rho=-2|D|^2({\rm Im}\Pi)$. Thus the only difference with the 
previous case is that the free photon propagator $D_0$ is replaced 
by the resummed photon propagator $D$, which includes the screening 
effects.

\newpage
\centerline{\bf FIGURE CAPTIONS}

\vspace*{0.5cm}
\noindent Figure 1: 
Left: Tree diagram for $e^-e^-$ scattering. 
Right: Two--loop self--energy diagram. The imaginary part, 
obtained by cutting the diagram through the fermion loop, corresponds
to the amplitude on the left, squared and integrated over phase space.

\vspace*{0.5cm}
\noindent Figure 2:
Left: Diagram for $e^-e^-$ scattering with a screened interaction. 
The resummed photon propagator is indicated by a blob. 
Middle: One--loop self--energy diagram with a resummed photon propagator. 
Right:
Emission of a virtual soft photon by a hard electron, corresponding to small 
angle $e^-e^-$ scattering.

\vspace*{0.5cm}
\noindent Figure 3: 
Tree diagrams for M\o ller scattering. 

\vspace*{0.5cm}
\noindent Figure 4: 
Two--loop self--energy diagrams corresponding to the scattering processes 
depicted 
in Fig.~3. The diagram on the left gives the direct and exchange contribution
while the diagram on the right is the interference term. 

\vspace*{0.5cm}
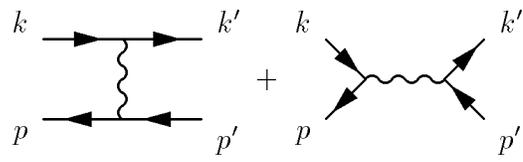
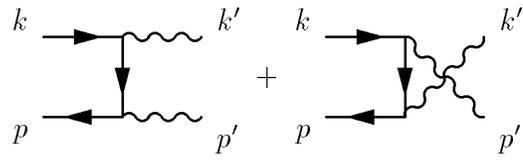
\noindent Figure 5:
Tree diagrams for (a) Bhabha scattering and (b) pair annihilation. 

\vspace*{0.5cm}
\noindent Figure 6: 
left: Tree diagram for $e^+e^-$ annihilation. As 
in Fig.~2 (left), the exchange diagram is negligible when the momentum 
carried by the internal propagator is soft.
Right:
Emission of a soft virtual positron (or absorption of a soft electron) 
by a hard positron, representing the 
contribution of soft momentum transfers to the process on the left. 

\vspace*{0.5cm}
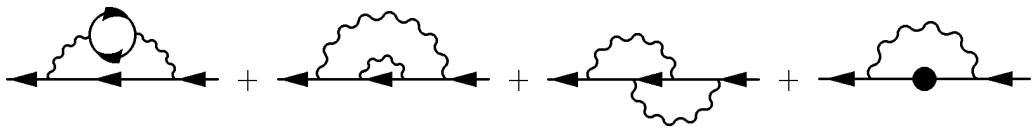
\noindent Figure 7:
Self--energy diagrams corresponding to positron scattering and annihilation. 
The three two--loop diagrams correspond to the hard contribution:
the first diagram corresponds to direct and exhange terms in 
Bhabha scattering (Fig.~5a), the second to direct and exchange terms in pair 
annihilation (Fig.~5b), while the third diagram, which can be cut in 
two different ways, gives the interference terms
of both processes; finally, the one--loop diagram on the right 
corresponds to the soft contribution to pair annihilation. 

\vspace*{0.5cm}
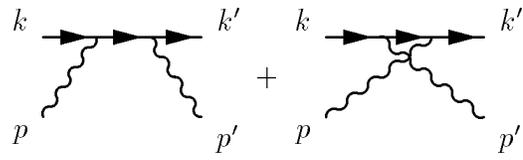
\noindent Figure 8: 
Tree diagrams for electromagnetic and Yukawa Compton scattering. 

\vspace*{0.5cm}
\noindent Figure 9: 
Left: Leading diagram for Compton scattering in the limit of 
soft exchanged momenta. 
Right: 
Emission of a soft virtual positron (or absorption of a soft electron) 
by a hard photon, corresponding to the process on the left. 

\vspace*{0.5cm}
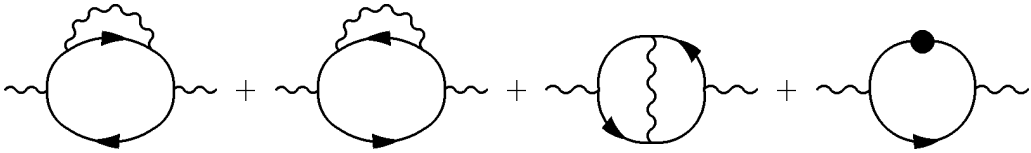
\noindent Figure 10:
Self--energy diagrams corresponding to Compton scattering. 
The three two--loop diagrams correspond to the hard contribution: 
the first and second diagrams correspond respectively to 
the first and second processes in Fig.~8, while the third diagram
is the interference term; finally, the one--loop diagram on the right 
corresponds to the soft contribution. 

\vspace*{0.5cm}
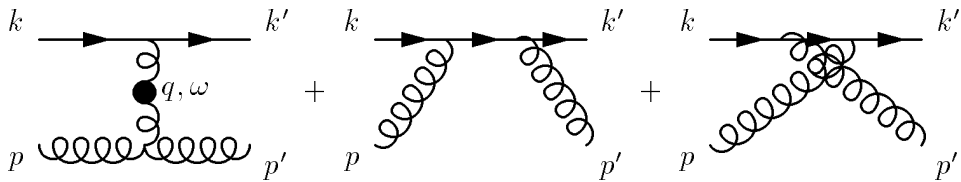
\noindent Figure 11:
Tree diagrams of gluon Compton scattering. The first of the three
diagrams dominates when the momentum carried by the internal 
gluon is soft. 

\vspace*{0.5cm}
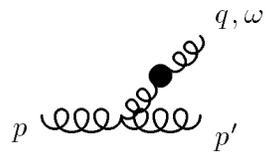
\noindent Figure 12:
Emission of a soft virtual gluon by a hard gluon, representing the
contribution of soft momentum transfers to the process depicted in 
Fig.~11. 

\vspace*{0.5cm}
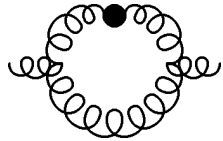
\noindent Figure 13: 
One--loop self--energy diagram contributing to the gluon damping rate 
through Compton scattering at small angles. 

\vspace*{0.5cm}
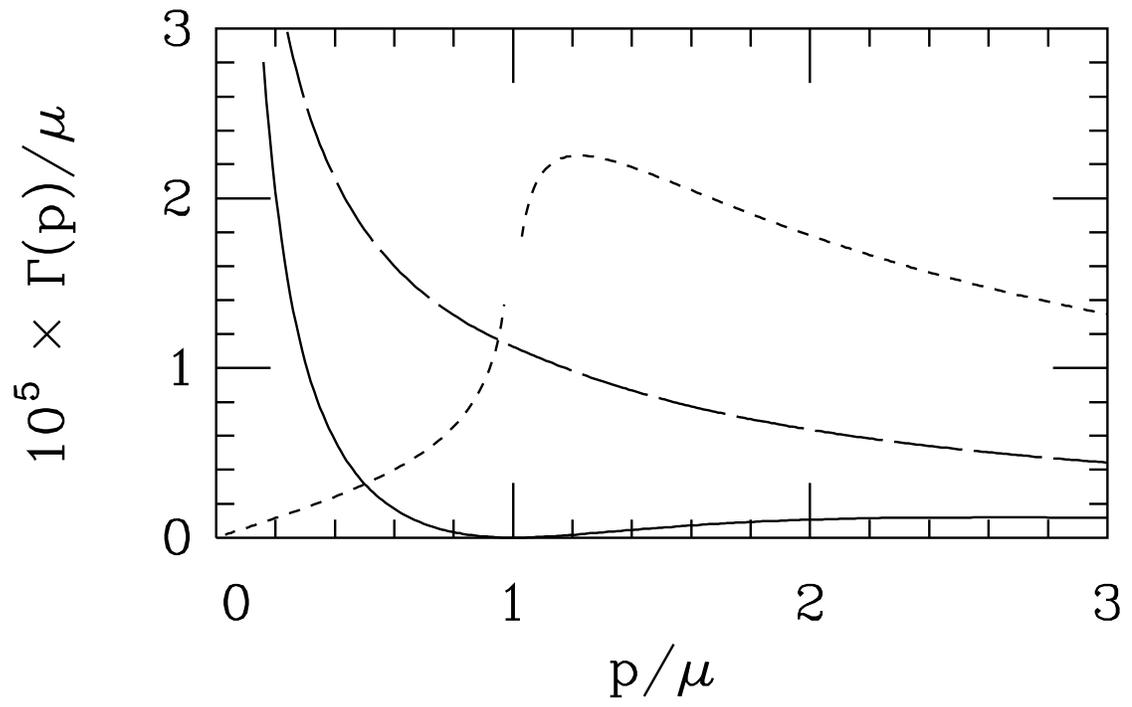
\noindent Figure 14:
Damping rate of hard one--particle excitations in the Yukawa theory as 
a function of their momentum $p$. 
We have chosen for $e$ the numerical value $e^2/4\pi=1/137$. 
Full line: fermion ($p>\mu$, eq.~(\ref{Gelyu>b})) 
and hole ($p<\mu$, eq.~(\ref{Gamyu<})) excitations; 
long dashes: antifermion (eqs.~(\ref{Gbhayu}) and (\ref{Gannyu})); 
short dashes: boson (eq.~(\ref{Ggamfinalyuka})). 

\vspace*{0.5cm}
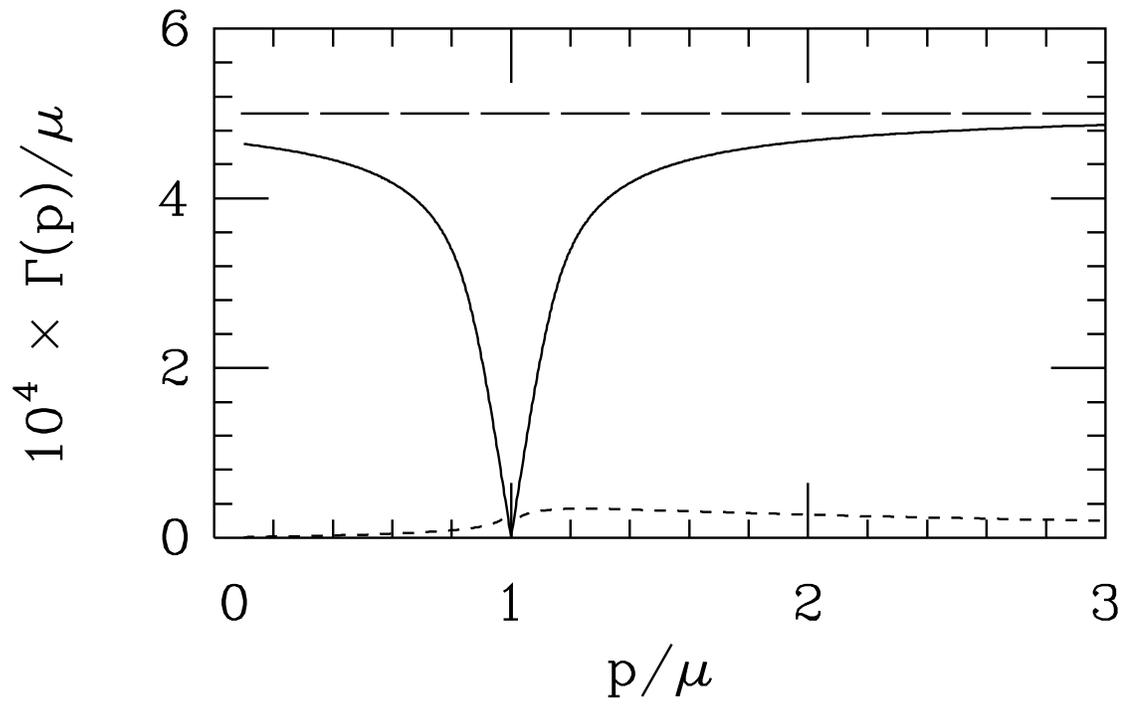
\noindent Figure 15:
Same as Figure 14 for QED interaction. 

\vspace*{0.5cm}
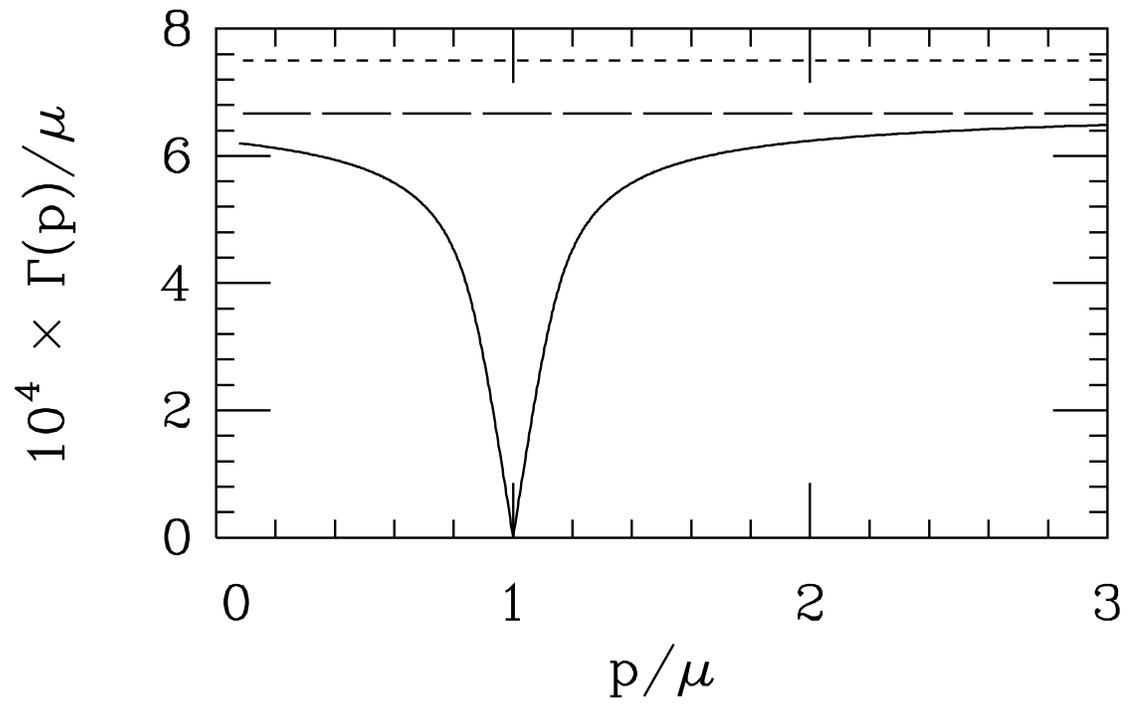
\noindent Figure 16:
Same as Figure 14 for QCD interaction, with $N_f=2$ flavors. 
The numerical value of the coupling constant is the same as in Figs. 14 
and 15. 

\vspace*{0.5cm}
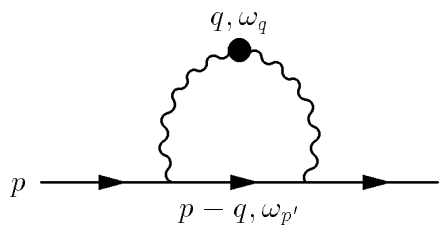
\noindent Figure 17: 
One loop resummed diagram contributing to the electron or quark 
damping rate. 

\end{document}